\documentclass[letterpaper,american,prr,twocolumn,amsmath,amssymb,superscriptaddress,nofootinbib]{revtex4-2}
\usepackage[T1]{fontenc}
\usepackage[active]{srcltx}
\usepackage{babel}
\usepackage{prettyref}
\usepackage{amsmath}
\usepackage{amssymb}
\usepackage{graphicx}
\usepackage[unicode=true,
 bookmarks=true,bookmarksnumbered=false,bookmarksopen=false,
 breaklinks=false,pdfborder={0 0 1},backref=false,colorlinks=false]
 {hyperref}
\hypersetup{pdftitle={Decomposing neural networks as mappings of correlation functions},
 pdfauthor={Kirsten Fischer, Alexandre Rene, Christian Keup, Moritz Layer, David Dahmen, Moritz Helias},
 colorlinks=true,linkcolor=black,citecolor=black,urlcolor=black,filecolor=black}

\makeatletter

\newcommand{\lyxdot}{.}

\usepackage[latin1]{inputenc}

\newrefformat{eq}{\hyperref[#1]{Eq.~(\ref{#1})}}
\newrefformat{cap}{\hyperref[#1]{Fig.~\ref{#1}}}
\newrefformat{fig}{\hyperref[#1]{Fig.~\ref{#1}}}
\newrefformat{tab}{\hyperref[#1]{Table ~\ref{#1}}}
\newrefformat{sec}{\hyperref[#1]{Section~\ref{#1}}}
\newrefformat{subsec}{\hyperref[#1]{Section~\ref{#1}}}
\newrefformat{cha}{\hyperref[#1]{Chapter~\ref{#1}}}
\newrefformat{app}{\hyperref[#1]{Appendix~\ref{#1}}}

\usepackage{MnSymbol}
\DeclareMathAlphabet\mathbb{U}{msb}{m}{n}

\usepackage{booktabs}% professional-quality tables
\usepackage[title]{appendix}

\usepackage[force]{feynmp-auto}

\makeatother

\begin{document}
 \renewcommand{\figurename}{FIG.} \renewcommand{\tablename}{TABLE} \renewcommand{\appendixname}{APPENDIX}
\global\long\def\llangle{\langle\!\langle}%
\global\long\def\rrangle{\rangle\!\rangle}%

\title{Neuronal architecture extracts statistical temporal patterns}
\author{Sandra Nestler}
\affiliation{Institute of Neuroscience and Medicine (INM-6) and Institute for Advanced
Simulation (IAS-6) and JARA-Institute Brain Structure-Function Relationships
(INM-10), Jülich Research Centre, Jülich, Germany}
\affiliation{RWTH Aachen University, Aachen, Germany}
\email{s.nestler@fz-juelich.de}

\author{Moritz Helias}
\affiliation{Institute of Neuroscience and Medicine (INM-6) and Institute for Advanced
Simulation (IAS-6) and JARA-Institute Brain Structure-Function Relationships
(INM-10), Jülich Research Centre, Jülich, Germany}
\affiliation{Department of Physics, Faculty 1, RWTH Aachen University, Aachen,
Germany}
\author{Matthieu Gilson}
\affiliation{Institute of Neuroscience and Medicine (INM-6) and Institute for Advanced
Simulation (IAS-6) and JARA-Institute Brain Structure-Function Relationships
(INM-10), Jülich Research Centre, Jülich, Germany}
\affiliation{Institut de Neurosciences des Systèmes (INS, UMR1106), INSERM-AMU,
Marseille (France)}
\date{\today}
\begin{abstract}
Neuronal systems need to process temporal signals. We here show how
higher-order temporal (co-)fluctuations can be employed to represent
and process information. Concretely, we demonstrate that a simple
biologically inspired feedforward neuronal model is able to extract
information from up to the third order cumulant to perform time series
classification. This model relies on a weighted linear summation of
synaptic inputs followed by a nonlinear gain function. Training both
-- the synaptic weights and the nonlinear gain function -- exposes
how the non-linearity allows for the transfer of higher order correlations
to the mean, which in turn enables the synergistic use of information
encoded in multiple cumulants to maximize the classification accuracy.
The approach is demonstrated both on a synthetic and on real world
datasets of multivariate time series. Moreover, we show that the biologically
inspired architecture makes better use of the number of trainable
parameters as compared to a classical machine-learning scheme. Our
findings emphasize the benefit of biological neuronal architectures,
paired with dedicated learning algorithms, for the processing of information
embedded in higher-order statistical cumulants of temporal (co-)fluctuations.
\end{abstract}
\maketitle

\section{Introduction\label{sec:introduction}}

It has long since been hypothesized that information about the environment
and internal states of humans and animals is represented in the correlated
neuronal activity. The most apparent examples include spike patterns
that are related to organization of cognitive motor processes and
visual perception \citep{Riehle97a,Kilavik09_12653,Siegle21_86,Shahidi19}.
Such patterns are also quantified indirectly via neuronal rhythms
\citep{Rubino-2006_1549,Nauhaus09_70,Sato12_218,fries15}. Furthermore,
the variability of network responses across trials when presenting
the same stimulus \citep{Arieli96} has been firstly shown to limit
encoding robustness, but was later found to serve a functional role
in a Bayesian context to represent (un)certainty about the presented
stimulus \citep{berkes2011spontaneous,orban2016neural}.

Structured variability is likely to play an important role in learning
by interacting with synaptic plasticity, in particular for models
like spike-timing-dependent plasticity (STDP) that are sensitive to
high-order statistics \citep{Gerstner96,Caporale2008,gjorgjieva2011triplet,gilson2011stdp}.
STDP implements an unsupervised learning rule similar to classical
Hebbian learning based on firing rates \citep{Hebb49,Hertz91} and
the BCM learning rule \citep{Bienenstock82,abraham2008metaplasticity,Izhikevich03}.
A counterpart for supervised learning has recently been proposed as
a basis for neuronal representations, which reconciles the apparent
contradiction between robust encoding by seemingly noisy signals \citep{Gilson20_1}.
This can be achieved by detecting and selecting correlated patterns
thanks to an adequate learning rule to update the synaptic weights
between neurons, thereby implementing a form of ``statistical processing''.
The focus on second-order statistics as a measure for spike trains
complements recent supervised learning schemes which either shape
the detailed spiking time generated by neurons or control their time-varying
firing rates \citep{Guetig06_420,ponulak10,Zenke18}.

An adjoint viewpoint to this biologically inspired learning is taken
up by machine learning. Though constructed from similar building blocks
(so called artificial neurons), the focus here lies in finding optimal
and efficient (re)encoding to process large amounts of data, the most
widespread application being classification. This field has produced
artificial neuronal architectures with impressive performance, even
better than human performance in the case of image recognition (see
\citep{Lecun2015deep,alzubaidi2021review} for reviews). However,
time series, which are naturally processed by biological neuronal
systems, can be challenging for these systems. For example, they are
often designed to operate on a feature space with fixed dimensionality
(including duration) as opposed to the learning of ongoing signals
that may have variable lengths. In this setting, artificial neural
networks are designed to account for different input durations across
samples. Recurrent neural networks, with their time-dependent network
states, handle these inputs by construction. Reservoir computing \citep{Jaeger01_echo,Maass02_2531},
which has been recently employed to capture statistical differences
in the input cumulants \citep{Nestler20_17380}, forms a straightforward
implementation of a trainable recurrent neural network. Also long-short
term memory (LSTM) units \citep{Hochreiter97} or the more recent
Gated Recurrent Units \citep{cho14properties} employ feedback connections
to process time series.

Time series processing has increasingly raised interest at the intersection
of biological learning and machine learning, based on a diversity
of network architectures and approaches \citep{Bagnall17,Ruiz21}.
Many studies rely on feedforward networks that have proven to be efficient
\citep{waibel1989phoneme,fuchs2009processing}; recurrent networks,
in contrast, are notoriously harder to train \citep{zenke15,hochreiter2001gradient},
despite progress for specific applications like time series completion
\citep{Sussillo09_544,DePasquale18}. Reservoir computing consists
in an intermediate approach, where a large (usually untrained) recurrent
network is combined with an easily trainable feedforward readout.
Its processing capabilities come from the combination of recurrent
connectivity with nonlinear units that performs complex operations
on the input signals, to be then selected by the readout to form the
desired output. The training of such feedforward readouts to capture
structured variability in a time series is our motivation and focus,
leaving out the reservoir here.

The present study considers statistical processing for the classification
of temporal signals by a feed-forward neuronal system, aiming to capture
structured variability embedded in time series. The goal is to automatically
select the relevant statistical orders, possibly combining them to
shape the output signal. We focus in this study on classification
that relies on the temporal mean of the output, which implies that
input cumulants at various orders need to be mapped to the first order
in output. The main inspiration is to design a biological setup consisting
of neurons that linearly sum input signals weighted by their afferent
connectivity weights, before passing the resulting signal to a non-linear
gain function. We also compare this biologically inspired architecture
with a machine learning approach, in terms of classification accuracy
and trainable weights (resources). \prettyref{fig:setup-biological}
displays a concise summary of the setup.

The remainder of this work is structured as follows: We first present
the models of neuronal classifiers with two flavors (\prettyref{sec:Model}):
a biological architecture and a machine learning architecture. We
show how they differ by the number of trainable weights (akin to resources)
and in their optimization. Then we describe the input time series
used to validate and compare these classifiers (\prettyref{sec:Cumulant-Generator}).
We rely on both synthetic data with controlled structures and contrast
between the classes as well as on real-world signals coming from \citealt{Chen15}.
Our results start with the synthetic datasets (\prettyref{sec:Training}),
to test whether the classifiers can extract statistics embedded in
time series that are relevant for the classification. In particular,
we examine whether the combination of different orders leads to synergistic
behavior, namely whether it increases the classification accuracy
(\prettyref{subsec:alphas}). Finally, we verify the practical applicability
of our framework to real-world datasets, comparing the performances
of both architectures to the state of the art (\prettyref{subsec:real-world-data}).

\section{Methods}

\subsection{Neuronal classifiers\label{sec:Model}}

\begin{figure}
\includegraphics[width=1\columnwidth]{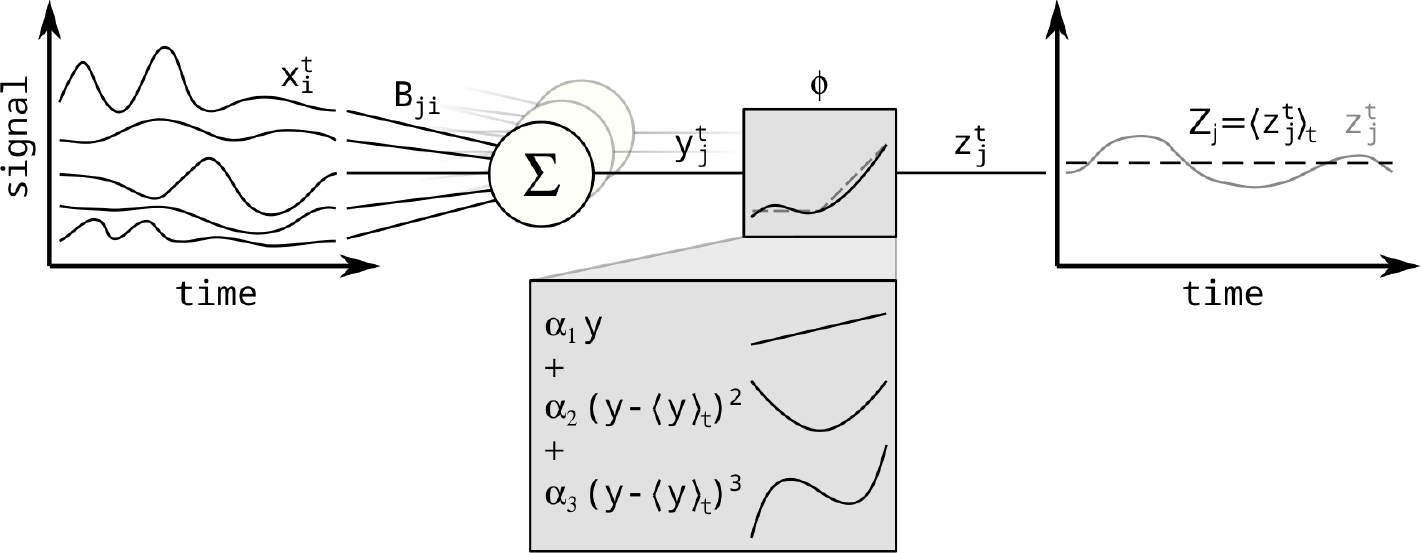}

\caption{\textbf{Biological interpretation of an order-selective perceptron.}
The input time series $x^{t}$ is linearly combined by the connectivity
matrix $B$ to form the intermediate variable $y^{t}=Bx^{t}$. The
latter is then passed on through a non-linear gain function $\phi$
to obtain the output $z^{t}$. The classification decision is made
with respect to the temporal average of $z^{t}$, which has the dimensionality
of the number of classes to predict, in a winner-take-all fashion.
The gain function $\phi$ is implemented by a third-order polynomial
with coefficients $\alpha_{1}$, $\alpha_{2}$, $\alpha_{3}$. The
contributions for orders two and three are the Taylor coefficients
of $\phi$ evaluated at the temporal mean of $\langle y^{t}\rangle_{t}$
over the duration of the input. In contrast, the linear contribution
directly acts on $z^{t}$. In the present figure, we consider $5$
input signals and $3$ output neurons, one of which is shown upfront.\label{fig:setup-biological}}
\end{figure}

We consider classification of multi-variate time series by a biologically
inspired neuronal architecture illustrated in \prettyref{fig:setup-biological}
that we term ``order selective perceptron'' (OSP). The multivariate
input time series $x_{i}^{t}$ with $1\le i\le N$ is linearly combined
via the trainable connectivity matrix $B_{ji}\in\mathbb{R}^{M\times N}$
to form the intermediate time-dependent variable $y_{j}^{t}=\sum_{i}B_{ji}x_{i}^{t}$.
Here $1\le j\le M$, where $M$ is the number of classes to be distinguished.
Then, a non-linear gain function $\phi$ is applied on $y_{j}^{t}$,
taken as a univariate signal, to obtain the output $z_{j}^{t}=\phi(y_{j}^{t})$.
Classification is performed based on the temporal mean (first-order
statistics) $Z_{j}=\langle z_{j}^{t}\rangle$. This gain function
$\phi$ is shaped to combine the cumulants of the intermediate variable
$y_{j}^{t}$, where the $n$-th statistical cumulant (evaluated over
time) of $y_{j}^{t}$ is denoted as $\langle\!\langle\big(y_{j}^{t}\big)^{n}\rangle\!\rangle_{t}=Y_{j}^{n}$
, resulting in $Z_{j}=\sum_{n}\alpha_{n}\langle\!\langle\big(y_{j}^{t}\big)^{n}\rangle\!\rangle_{t}=\sum_{n}\alpha_{n}Y_{j}^{n}$.
The $n$-th cumulant $Y_{j}^{n}$ reflects the corresponding cumulants
at the same order for the multivariate inputs $x_{i}^{t}$ as 
\begin{equation}
Y_{j}^{n}=\langle\!\langle\big(y_{j}^{t}\big)^{n}\rangle\!\rangle_{t}=\sum_{i_{1},\ldots,i_{n}=1}^{N}B_{ji_{1}}\cdots B_{ji_{n}}\langle\!\langle x_{i_{1}}^{t}\cdots x_{i_{n}}^{t}\rangle\!\rangle_{t}\,.\label{eq:cumulant-relation}
\end{equation}
Thus, a general gain function combines several statistical orders
of the input to generate the mean $Z_{j}$ of the output activity.
Here, we aim to perform the classification based on the first three
cumulants. The gain function is therefore implemented as a trainable
third-order polynomial of the intermediate variable $y_{j}^{t}$,
see the inset panel in \prettyref{fig:setup-biological}.

\subsubsection{Link to covariance perceptron}

\begin{widetext}

\begin{figure}
\begin{minipage}[t][1\totalheight][c]{0.45\columnwidth}%
\includegraphics[width=0.9\columnwidth]{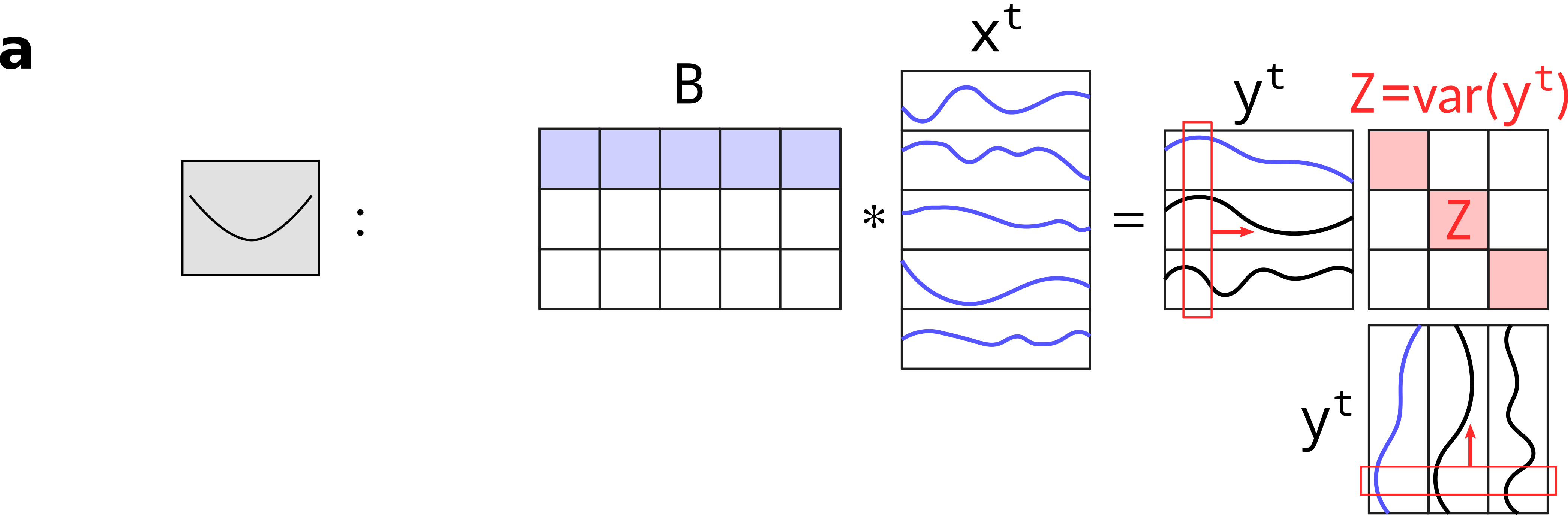}

\medskip{}

\includegraphics[width=0.9\columnwidth]{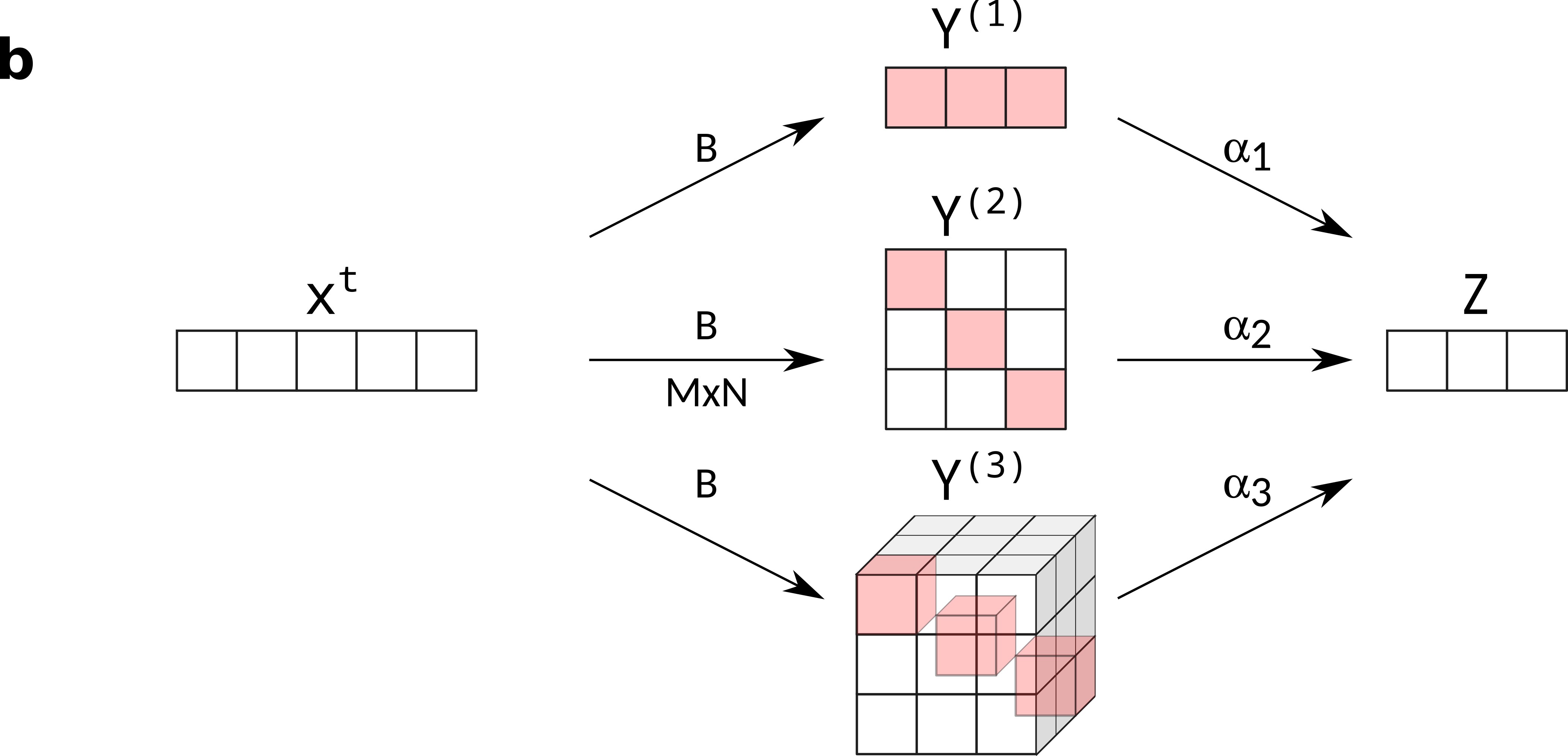}%
\end{minipage}%
\begin{minipage}[t][1\totalheight][c]{0.45\columnwidth}%
\includegraphics[width=0.9\columnwidth]{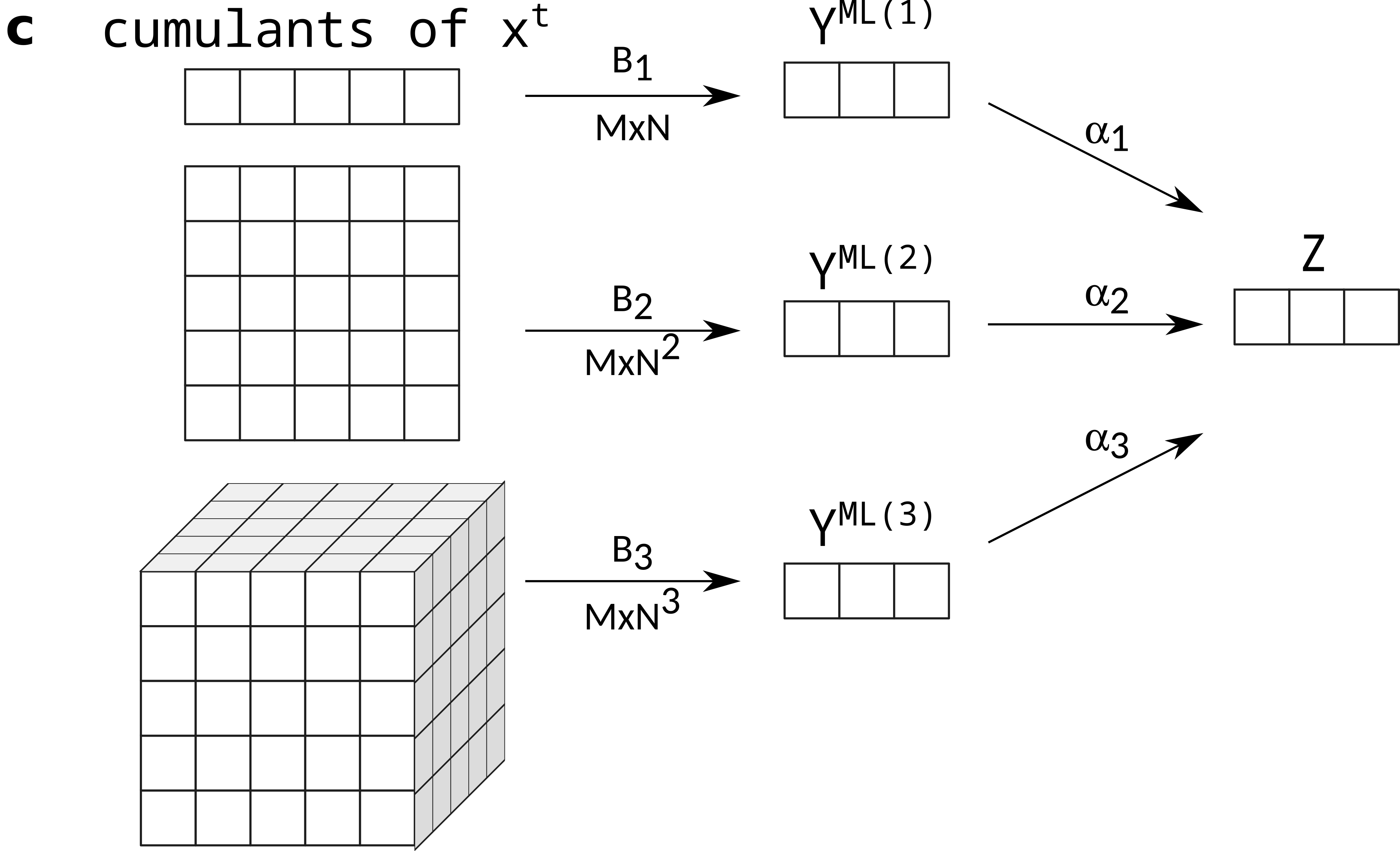}%
\end{minipage}

\caption{\textbf{Extraction of cumulants from a time series. (a) Covariance
perceptron.} The classification of the centered time series $x^{t}$
is based on its covariance structure. The output $Z=(Z_{1},\ldots,Z_{M})$
has dimension equal to the number of classes $M$. For each class
$j$ ($1\le j\le M)$, the covariance perceptron calculates an intermediate
variable $y_{j}^{t}=\sum_{i}B_{ji}x_{i}^{t}$ that is then passed
though a quadratic function $z_{j}^{t}=(y_{j}^{t})^{2}$. The temporal
mean $Z_{j}=\langle z_{j}^{t}\rangle=\langle\!\langle(y_{j}^{t})^{2}\rangle\!\rangle_{t}$,
which therefore equals the variance of $y^{t}$ is used for classification
in a winner-take-all manner. This is equivalent to retaining the diagonal
of the covariance matrix of $y_{1\le j\le M}^{t}$. \textbf{(b) Statistical
processing for order selective perceptron (OSP) model.} Similar to
panel (a), the effect of the gain function $\phi$ in \prettyref{fig:setup-biological}
can be represented by an individual path for each cumulant $\langle\!\langle\big(y_{j}^{t}\big)^{n}\rangle\!\rangle_{t}$
of $y_{j}^{t}=\sum_{i}B_{ji}x_{i}^{t}$. The cumulants of $y^{t}$
are thus given by the cumulants of the inputs $x^{t}$ of the same
order $n$ (\prettyref{eq:cumulant-relation}). The cumulant of order
$n$ can be seen as an outer product with dimensionality $M^{n}$:The
mean is represented by a line (vector), the covariance by a square
(matrix; as in panel a) and the third order cumulant by a cube (third
order tensor). The effect of applying the polynomial $\phi(y)=\alpha_{1}y+\alpha_{2}\tilde{y}^{2}+\alpha_{3}\tilde{y}^{3}$
with $\tilde{y}^{t}=y^{t}-\langle y^{t}\rangle_{t}$ and then taking
the mean to obtain $Z$ is to combine these contributions linearly
with a weight vector $\alpha=(\alpha_{1},\alpha_{2},\alpha_{3})$
. \textbf{(c) Machine-learning (ML) model.} Statistical processing
can alternatively be implemented by a network that is fed by the input
cumulants directly. For each order, a linear layer is set up with
the corresponding dimensionality: the total size of the input cumulant
multiplied by the number of outputs (i.e. of classes). In a second
layer, a linear combination of the outputs for the individual orders
is trained.\label{fig:setup-schematic}}
\end{figure}

\end{widetext}

Before explaining the OSP in more detail, we briefly present its operational
link with the covariance perceptron \citep{Gilson20_1,Dahmen20_354002}.
The goal of the latter is to operate on the second-order statistics
embedded in the time series, instead of the first-order statistics
that is equivalent to the classical perceptron applied to mean activity
\citep{Rosenblatt58,Minsky69,Widrow60_96}. It can be formalized as
in \prettyref{fig:setup-schematic} (a), where the linear mixing is
based on the weight matrix $B_{ji}^{\text{P}}$ applied to the $N$-dimensional
time-dependent inputs $x_{i}^{t}$ (after demeaning) gives the $M$-dimensional
intermediate variable $y_{j}^{t}$ 
\begin{equation}
y_{j}^{t}=\sum_{i}B_{ji}^{\text{P}}\left(x_{i}^{t}-\langle x_{i}^{t}\rangle_{t}\right)\,.
\end{equation}
From the outer product of $y_{j}^{t}$ that forms the covariance matrix,
the classification only considers the variances (diagonal matrix elements)
defined as
\begin{align}
Z_{j} & =\langle\!\langle(y_{j}^{t})^{2}\rangle\!\rangle_{t}\\
 & =[\text{diag}(B^{\text{P}}\Sigma B^{\text{P}\,\mathrm{T}})]_{j}\,.
\end{align}
 Thanks to the quadratic operation, the mean value of $Z_{j}$ depends
on the input covariances $\Sigma_{ii^{\prime}}=\langle\!\langle x_{i}^{t}x_{i^{\prime}}^{t}\rangle\!\rangle_{t}$
for all possible indices $i$ and $i^{\prime}$. In this scheme, there
are only $M\cdot N$ weights in matrix $B^{\text{P}}$ that can be
trained. Thus the number of parameters can be compared to a machine-learning
approach that applies a perceptron the entries of the covariance matrix
$Z_{j}^{\prime}=\sum_{ii^{\prime}}B_{jii^{\prime}}^{\text{ML}}\Sigma_{ii^{\prime}}$,
which involves $M\cdot N^{2}$ trainable parameters of the tensor
$B^{\text{ML}}$.

\subsubsection{Order selective perceptron (OSP)\label{subsec:Order-selective-perceptron}}

Now we consider the order selective perceptron (OSP), a feedforward
network that extends the covariance perceptron to combine the first
three statistical orders of the input time series $x^{t}$ via the
calculation of the output time series $z^{t}$, illustrated in \prettyref{fig:setup-biological},
as

\begin{align}
y_{j}^{t} & =\sum_{i}B_{ji}x_{i}^{t}\,,\label{eq:input-layer}\\
z^{t} & =\phi(y^{t})\:=\alpha_{1}y+\alpha_{2}\tilde{y}^{2}+\alpha_{3}\tilde{y}^{3}.\label{eq:output-layer}
\end{align}
Here the linear weights $B_{ji}$ are trainable, as well as the coefficients
$\alpha_{1\le i\le3}$ of the (nonlinear) polynomial gain function
$\phi$. Note that $\tilde{y}^{t}=y^{t}-\langle y^{t}\rangle_{t}$
is the demeaned version of time series $y^{t}$.

The key of the statistical processing is the following: the cumulants
at orders 1 to 3 can be calculated using the polynomial of order 3
applied to the intermediate variable $y_{j}^{t}$, as the second and
third cumulants are the central moments (including the demeaning).
An abstract representation of the resulting computation is illustrated
in \prettyref{fig:setup-schematic} (b). Each path of the network
corresponding to a specific cumulant of order $n$ of the intermediate
variable $y^{t}$. The OSP computation can thus be understood via
the cumulants of order $n\in\{1,2,3\}$ of $y_{j}^{t}$ as illustrated
in \prettyref{fig:setup-schematic}(b) 
\begin{align}
Y_{j}^{(\!n\!)} & =\langle\!\langle(y_{j}^{t})^{n}\rangle\!\rangle_{t}\,,\label{eq:cumulant-layer}
\end{align}
where the symbol $\langle\!\langle(y_{j}^{t})^{n}\rangle\!\rangle_{t}$
stands for the mean over time $\langle y_{j}^{t}\rangle_{t}$ for
$n=1$, and the $n$-th power of the demeaned variable $\langle(y_{j}^{t}-\langle y_{j}^{t}\rangle_{t})^{n}\rangle_{t}=\langle(\tilde{y}^{t})^{n}\rangle_{t}$
for $n\in\{2,3\}$; they correspond to the red diagonal matrix or
tensor elements in \prettyref{fig:setup-schematic}(b). Importantly,
each cumulant $Y_{j}^{(\!n\!)}$ for order $n$ depends on the $n$-th
statistical order of $x$ (see also \prettyref{eq:cumulant-relation}).

Note that it is straightforward to generalize this scheme to arbitrary
cumulant orders, beyond the first three orders considered here, because
a cumulant of any order $n$ can be expressed as a linear combination
of moments of orders $1,\ldots,n$; beyond third order, however, they
cease to be identical to the $n$-th centered moments \citep{Gardiner85}.
The resulting computation of the OSP is thus the combination in the
output mean $Z_{j}=\langle z_{j}^{t}\rangle_{t}$ of the different
cumulant orders $n$
\begin{equation}
Z_{j}=\sum_{n}\alpha_{n}Y_{j}^{(\!n\!)}\,,\label{eq:order-select-layer}
\end{equation}
which is the basis for the classification via
\begin{equation}
\arg\max(Z_{j}-\theta_{j})
\end{equation}
with some trainable thresholds $\theta_{j}$. The key point here is
that the selection of the informative cumulant orders for the classification
is managed by tuning the parameters $\alpha_{n}$.

We now want to evaluate how the OSP combines different cumulant orders
to perform classification. From the network after training, which
we refer to as the ``full'' model with all coefficients $\alpha_{n}$,
we can ignore contributions from given cumulant orders by setting
the corresponding parameter $\alpha_{n}$ to zero. In this way, we
create a ``single order'' OSP for order $n$ with $\alpha_{n^{\prime}}=0\quad\forall n{}^{\prime}\neq n$.
This network is then classifying based on the $n$-th cumulant only
with the corresponding learned parameter $\alpha_{n}$. With this
approach we quantify the contribution of an individual statistical
order $n$ to classification in the full model.

\subsubsection{Machine-learning classifier\label{subsec:Machine-learning-classifier}}

For comparison, we consider a machine learning (ML) architecture that
computes the same statistical cumulants, as features for a linear
perceptron. The operational difference is that the weights directly
map all features to the next layer, as displayed in \prettyref{fig:setup-schematic}
(c)
\begin{align}
Y_{j}^{\text{ML}(\!n\!)} & =\sum_{i_{1}\ldots i_{n}}B_{ji_{1}\ldots i_{i}}^{\text{ML}(\!n\!)}\langle\!\langle x_{i_{1}}^{t}\cdots x_{i_{n}}^{t}\rangle\!\rangle_{t}^{n}\,,
\end{align}
yielding a number of trainable parameters equal to $MN^{n}$ for order
$n$. As before, the different orders are combined and selected using
\prettyref{eq:order-select-layer}. Note that the ML architecture
differs from the OSP in that it outputs scalar values, not a time
series. To be able to further compare this ML architecture to the
OSP, we randomly select $MN$ trainable weights from the whole $MN^{n}$
entries of $B^{\mathrm{ML}}$ (the other weights being fixed to random
values) to build the 'constrained ML' architecture which has the same
number of trainable parameters as the OSP. This way, the complexity
of the two systems is comparable.

\subsubsection{Training process\label{subsec:Training-process}}

Training of both network types is conducted via a stochastic gradient
descent using a scaled squared error loss $\varepsilon$ for each
data sample $\nu$ of the form
\begin{equation}
\varepsilon(x^{\nu})\sim\frac{1}{2}\sum_{j=1}^{M}\left(\bar{Z}_{j}^{c(\nu)}-(Z_{j}^{\nu}-\theta_{j})\right)^{2}\,,
\end{equation}
where $\bar{Z}_{j}^{c(\nu)}=\delta_{j,c(\nu)}$ is the one-hot encoded
target for the mean, with $c(\nu)\in\{1,\ldots,M\}$ the ground truth
of the class of data sample $\nu$ and $\theta_{j}$ is a trainable
bias. In the binary classification problem, the one-hot encoded network
output has dimension $M=2$. In addition, we add an L2 regularization
term $\mu\,\Vert\alpha\Vert^{2}$ to the loss with a fixed, small
$\mu$ to decrease the entries of the order selection parameter $\alpha$.
This ensures that uninformative layers lead to a vanishing contribution
to $\alpha$, but also implements a winner-take-all mechanism when
there is a strong imbalance between the difficulty of classification
for each individual order. Non-vanishing entries in $\alpha$ hence
indicate the contribution of the corresponding statistical order to
the classification decision. We therefore quantify the synergy of
the OSP by the difference in performance of the full OSP model with
models that were pruned after training such that they only use a single
order $n$ at a time. This is achieved by setting the parameters $\alpha_{n^{\prime}\neq n}=0$
for the other orders. The participation ratio $\rho$,
\begin{equation}
\rho(\alpha)=\frac{\sum_{n}|\alpha_{n}|}{\left(\sum_{n}|\alpha_{n}|^{2}\right)^{\frac{1}{2}}}\,,\label{eq:participation-ratio}
\end{equation}
can be used as a measure of sparsity to relate the synergy to how
many relevant statistical orders the network found in the data. It
is minimal when there is only a single non-zero $\alpha_{n}$ and
maximal when all $\alpha_{n}$ are equal. Thus, it increases the more
different orders $n$ are combined.

\subsection{Synthetic datasets with controlled cumulant structure\label{sec:Cumulant-Generator}}

We design synthetic datasets to assess the ability of the proposed
networks to capture specific statistical orders in the input signals.
To that end, we generate two groups of multivariate time series with
desired statistics up to the third order. We control how the two groups
differ with regard to the first three cumulants. Concretely, the time
series are generated by simulating in discrete time a process defined
by a stochastic differential equation (SDE), whose samples in the
infinite time limit follow a Boltzmann probability distribution
\begin{equation}
p(x)\sim\exp(-\beta L[x]).\label{eq:probability-density}
\end{equation}
Samples $x(t)=\{x_{i}(t)\}$ are produced by the following process
defined with $L[x]$ as Lagrangian 
\begin{equation}
\frac{\partial x_{i}(t)}{\partial t}=-\Gamma\frac{\partial L[x]}{\partial x_{i}}(t)+\xi_{i}(t)\,.\label{eq:generative_model}
\end{equation}
Here, $\xi$ is the stochastic Gaussian increment (or Wiener process
or more colloquially white noise) that obeys $\langle\xi\rangle=0$,
$\langle\!\langle\xi_{i}(t),\xi_{j}(t^{\prime})\rangle\!\rangle=D\delta_{ij}\delta(t-t^{\prime})$
and $\Gamma$ is a constant parameter. Using a fluctuation-dissipation
theorem \citep{Goldenfeld92} one finds that $\beta=2\frac{\Gamma}{D}$
in order for $x$ to follow the distribution \eqref{eq:probability-density}
shaped by the choice of $L[x]$.

The particular case of a Gaussian distribution corresponds to a quadratic
Lagrangian 
\begin{equation}
L[x]=m^{T}x+\frac{1}{2}x^{T}Jx\,,
\end{equation}
where $J=J^{\mathrm{T}}$ is a symmetric matrix, such that the mean
$\mu=\langle x\rangle$ and covariance $\Sigma_{ij}=\langle\!\langle x_{i}x_{j}\rangle\!\rangle$
of the distribution read 
\begin{align}
\mu & =-J^{-1}m\,,\\
\Sigma & =(\beta J)^{-1}\,,
\end{align}
while the third order, $S_{ijk}=\langle\!\langle x_{i}x_{j}x_{k}\rangle\!\rangle$,
vanishes. The more general case with the additional 3rd order statistics
can then be developed in the spirit of field theory with a small perturbation
(or correction) on the Gaussian case. The corresponding Lagrangian
reads
\begin{equation}
L[x]=m^{T}x+\frac{1}{2}x^{T}Jx+\frac{1}{3!}\sum_{ijk}K_{ijk}x_{i}x_{j}x_{k}.\label{eq:third_order_Lagrangian}
\end{equation}
For a sufficiently small tensor $K$, the first three cumulants can
be approximated as (See appendix, \prettyref{sec:Compute-statistics})
\begin{align}
\mu_{i} & \approx-\sum_{j}(J^{-1})_{ij}m_{j}-\frac{1}{2\beta}\sum_{jkl}(J^{-1})_{ij}K_{jkl}(J^{-1})_{kl}\,,\\
\Sigma_{ij} & \approx\frac{1}{\beta}(J^{-1})_{ij}\,,\\
S_{ijk} & \approx-\frac{1}{\beta^{2}}\sum_{i^{\prime}j^{\prime}k^{\prime}}K_{i^{\prime}j^{\prime}k^{\prime}}(J^{-1})_{ii^{\prime}}(J^{-1})_{jj^{\prime}}(J^{-1})_{kk^{\prime}}\,.
\end{align}
This approximation only holds for a small deviation from the Gaussian
case, which limits the amplitude of the third order statistics. Concretely,
we introduce a safety parameter $s$ to ensure that a local minimum
of the potential $L[x]$ exists and that the distance between the
local minimum and the next maximum is large enough to fit $s$ standard
deviations of $x$ in any direction. When integrating the SDE, large
$s$ ensures a low probability for a sample initialized in that local
minimum to evolve further than the adjacent local maximum and prevents
it of falling into the unstable region. In this case, the ratio of
third to second order cumulants of $x$, projected to any eigendirection
$e^{(\!v\!)}$ of $J$, is maximally allowed to be 
\begin{equation}
\left|\frac{\sum_{ijk}S_{ijk}e_{i}^{(\!v\!)}e_{j}^{(\!v\!)}e_{k}^{(\!v\!)}}{\left(\sum_{ij}\Sigma_{ij}e_{i}^{(\!v\!)}e_{j}^{(\!v\!)}\right)^{\frac{3}{2}}}\right|=\frac{1}{s}\,.
\end{equation}
In the appendix, \prettyref{sec:Select-K}, we show how to compute
a suitable tensor $K$ under this constraint. Additionally, we redraw
samples that deviate too strongly from the local minimum of $L[x]$,
which corresponds to introducing an infinite potential wall where
$L[x]$ negatively exceeds the local minimum. Due to the low probability
mass in that area for sufficient safety $s$, we do not need to account
for this potential wall in the cumulant estimates. This way, we avoid
issues with the probability density \prettyref{eq:probability-density}
not being normalizable. Alternatively, it would also be possible to
include a positive definite fourth order term in the Lagrangian and
account for this in the calculation of the arising data statistics,
however the more simple third order with redrawing and a safety parameter
$s=5$ proved to work satisfactorily for the purposes of this work.

We create datasets where the classification difficulty is controlled
for each statistical order. To do so, we generate time series grouped
in two equally-sized classes, whose cumulants differ by a scaling
factor that serves as contrast. For example, we draw a reference class
with some randomly drawn cumulants. A class contrast of $1.5$ in
the mean and $1$ otherwise then corresponds to the second class having
$1.5$ times larger mean and all other cumulants exactly the same.
Due to the stability constraint, the third order cumulant may not
become arbitrarily large. Practically, we use the class with the larger
third order as the fixed cumulant and generate the second class from
the first using the reciprocal contrast. This way, in any order, the
larger cumulant is the contrast times the smaller cumulant.

\subsection{Real datasets}

We test our model's ability to classify on a selection of time-series
classification datasets that have previously been used as benchmarks
\citep{Chen15}. We select a subset of $18$ multivariate time series
datasets of which the data dimensionality allows us to compute the
cumulants of the classes directly on the input. Before training, we
shift and rescale the data to centralize them around zero and obtain
unit variance over all data points. This is required because of the
unboundedness of the gain function, which is a third order polynomial,
as well as numerical stability when computing the cumulants on the
data directly.

\section{Results}

We first benchmark the models proposed in \prettyref{sec:Model} with
synthetic data generated with controlled statistical patterns (see
\prettyref{sec:Cumulant-Generator}). We show how the trained parameters
can be analyzed to infer what statistical orders are relevant for
the classification. The trained network parameters $\alpha_{1\le l\le3}$
in particular allow us to determine the optimal gain function for
each dataset. Subsequently we study classification of real multivariate
time series. We find that the OSP classifies efficiently compared
to the ML model and we observe that the various datasets combine the
statistical orders in a different way. The optimal network typically
operates with a synergy of different orders.

\subsection{Benchmarking with synthetic data\label{sec:Training}}

Classification may be performed based on one or several statistical
orders measured on each sample time series. Each order thus defines
a contrast between the two classes, which corresponds to a baseline
classification accuracy when using the corresponding statistical order
alone to perform the classification. This contrast is controlled by
the coefficients of the two classes in the generative model, as explained
in \prettyref{eq:generative_model} and \prettyref{eq:third_order_Lagrangian}.

We also compare the respective classification accuracy of the biological
and machine learning architectures while varying the class contrasts
to see how they are combined. Note that cumulants of orders higher
than three are non-vanishing and may also hold information on the
classes, but the neuronal networks by construction neglect them. Each
of the two classes comprises $100$ samples, which have dimensionality
$N=5$ and $T=100$ time steps per sample. We repeat the classification
$25$ times. For each binary classification, two of the three contrasts
are chosen to be different from $1$, thus contain the information
about the class, while the third contrast is the same for both classes,
having contrast of unity.

\begin{widetext}

\begin{figure}
\begin{centering}
\includegraphics[width=1\textwidth]{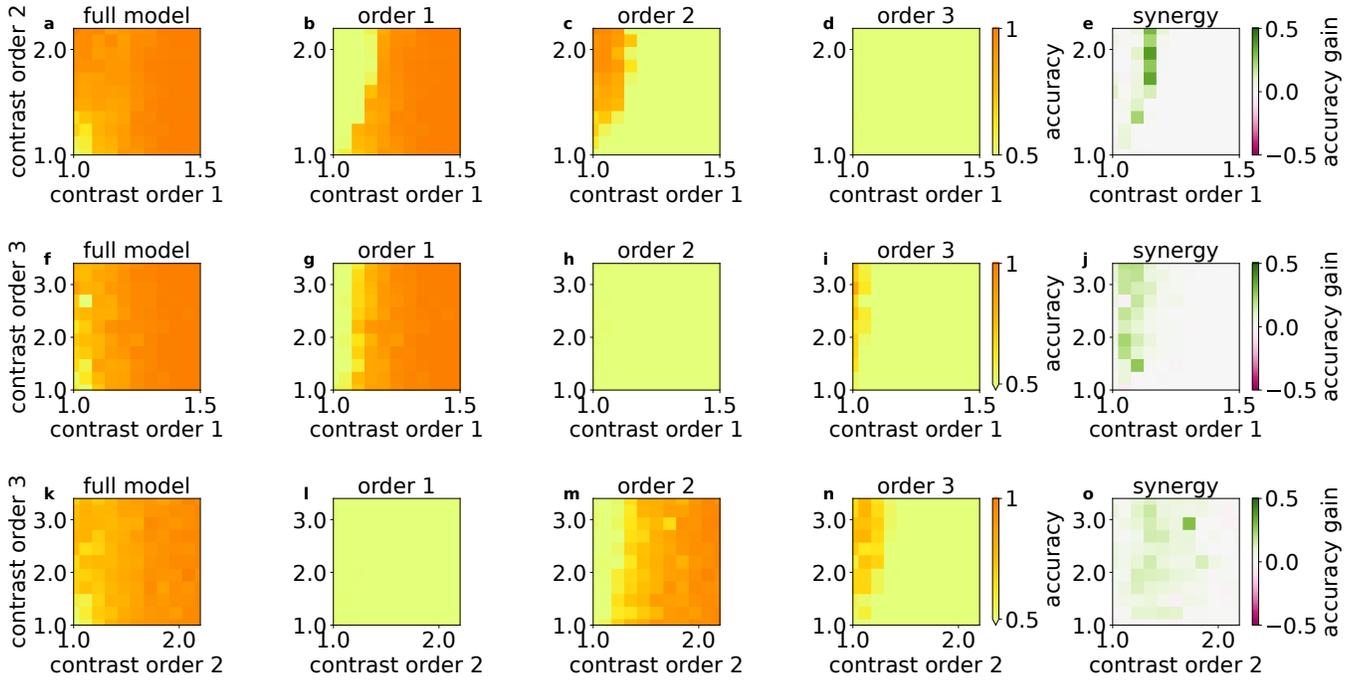}
\par\end{centering}
\caption{\textbf{Classification accuracies for varying task difficulties.}
Datasets with two equal-sized classes that differ in two statistical
orders are classified using the OSP. Along the axes of each diagram
the contrast between the two classes is varied between one (no class
difference) to an arbitrarily chosen upper scale.\textbf{ (a, f, k)}
Test set accuracy of the full model. \textbf{(b, g, l)} Accuracy obtained
by the OSP when pruning all $\alpha_{n}$ except $\alpha_{1}$ after
training. \textbf{(c}, \textbf{d, h, i, m, n)} Analogous to (b) for
$\alpha_{2},\alpha_{3}$, respectively. \textbf{(e, j, o)} Synergy,
which is the difference in accuracy between the full model and the
best single order. In (a - e), classes are separated by a difference
in the mean and covariance. In (f - j), classes are separated by a
difference in the mean and third order cumulant. In (k - o), classes
are separated by a difference in the covariance and third order cumulant.
\label{fig:Acc-synthetic}}
\end{figure}

\end{widetext}

We start with testing whether the OSP can be efficiently trained to
perform classification. \prettyref{fig:Acc-synthetic} shows the training
accuracy for (Gaussian) inputs where the contrasts of orders one and
two are varied, while the third cumulant is zero. Accuracy ranges
from chance level ($50\%$ for the considered case of two balanced
classes, in yellow) to perfect discrimination corresponding to $100\%$
(in orange). Here, successful training means that the OSP captures
the most relevant class contrast(s) among the three cumulant orders.
As expected, the accuracy of the full model increases when either
contrast of order one or two increases (see \prettyref{fig:Acc-synthetic}
(a)).

Investigating the contribution of individual orders in isolation,
the accuracy increases with the corresponding contrast between the
classes, as shown when varying either order one or two in \prettyref{fig:Acc-synthetic}(b,
c). The contribution of the uninformative third order stays chance
level \prettyref{fig:Acc-synthetic}(d). In \prettyref{fig:Acc-synthetic}
(f - o), the same qualitative behavior is observed for all different
combinations of informative and uninformative orders. This shows that
the OSP successfully manages to capture the relevant orders for classification,
leading to the question of their combination when two or more orders
are relevant.

\subsubsection{Synergy between relevant cumulant orders}

The accuracy of the full model (\prettyref{fig:Acc-synthetic} (a))
is larger than the accuracy obtained when single orders are taken
individually (\prettyref{fig:Acc-synthetic} (b-d)). The difference
between the accuracy for full training and the maximum of single-order
accuracies, shown in \prettyref{fig:Acc-synthetic} (e), can be used
as a measure of synergy: it indicates how the OSP makes use of the
combination of the different statistical orders, specifically in the
border between the regions of high single-order accuracy. Thus, the
OSP (full model) combines the different orders relevant for the classification,
beyond simply selecting the most informative order.

We observe a bias between the accuracies for single-order models.
Although the synthetic data were generated to have comparable contrasts
across the orders, we see that the border where the OSP (full model)
switches from one order to another order does not follow a diagonal
with equal contrast. This indicates a bias favoring lower statistical
orders over higher ones. This may partly come from the fact that higher
statistical orders are noisier because they require more time points
for estimation. The estimation error causes fluctuations which may
affect the training by the parameter updates. In addition to this
implicit bias towards extraction of information from lower orders,
the normalization imposed on the $\alpha_{n}$ during training (see
\prettyref{subsec:Training-process} for details) implements a soft
winner-take-all mechanism that likely reinforces the bias(es). In
\prettyref{sec:other-synergy}, we discuss the accuracy gain of the
OSP compared to a restricted OSP, as opposed to the pruned OSP, which
can be seen as an alternative way to quantify synergy.

\subsubsection{Comparison with machine learning architecture}

\begin{figure}
\begin{centering}
\includegraphics[width=1\columnwidth]{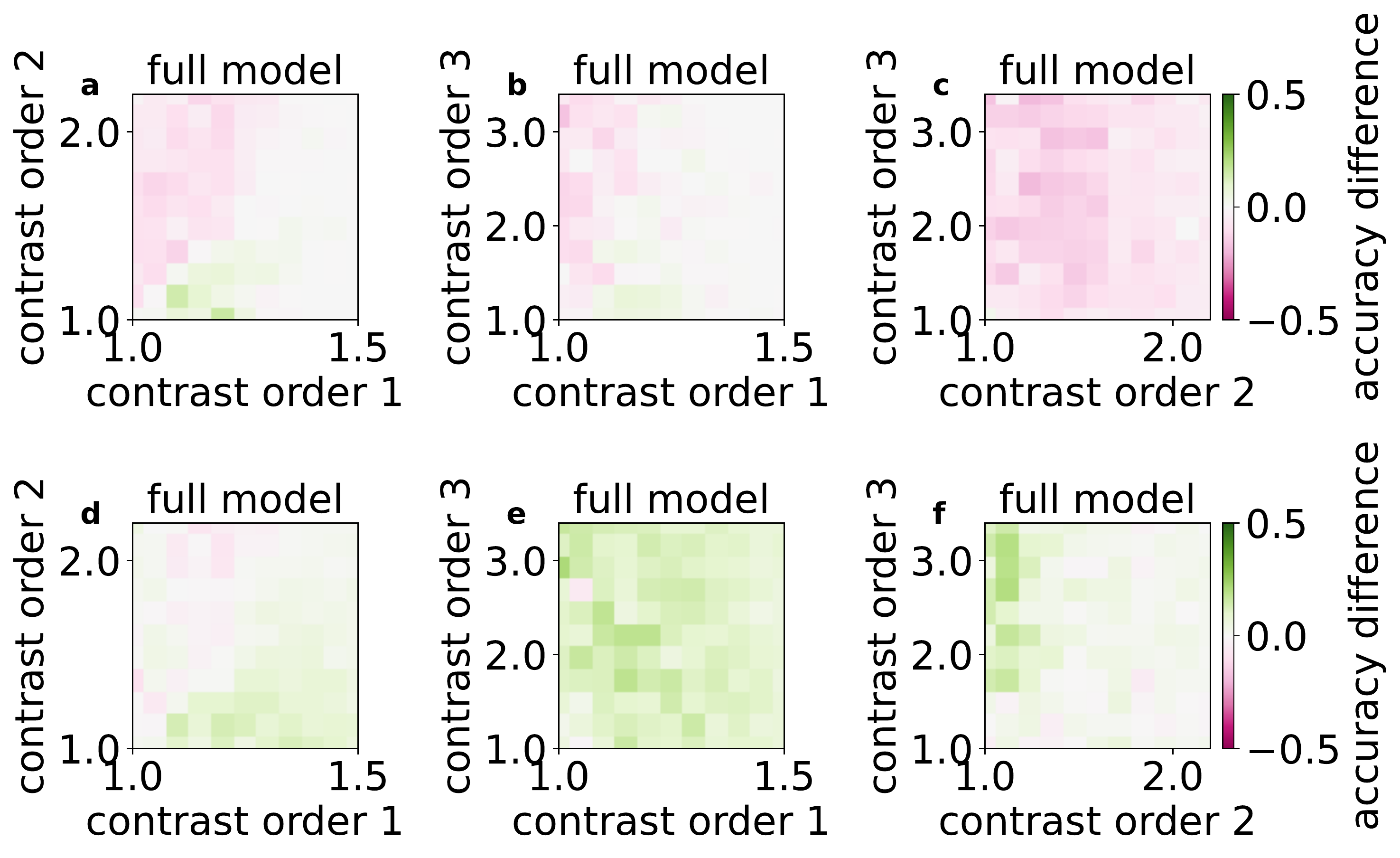}
\par\end{centering}
\caption{\textbf{Comparison of the OSP and the ML model.} Difference in accuracy
between the OSP and the ML model for the datasets in \prettyref{fig:Acc-synthetic}.
\textbf{(a - c)} Accuracy comparison between the ML model and the
OSP. \textbf{(d - f)} Accuracy comparison between the constrained
ML model and the OSP. The color code shows the difference in accuracy
between the respective ML model and the full OSP model. \label{fig:OSP-vs-ML}}
\end{figure}

Last, we compare the OSP to the ML model that has more trainable weights
(or resources). As shown in Appendix \prettyref{sec:Artificial-ML},
the ML network performs well and can discriminate between informative
orders. From \prettyref{fig:OSP-vs-ML} (a - c), we find that the
accuracy for the OSP is similar to that of the ML, although mostly
a bit lower presumably due to its lower number of trainable parameters
(weights $B$, compare \prettyref{fig:setup-schematic}(b-c)). Nevertheless,
it can efficiently combine ``information'' distributed across statistical
orders to perform the classification, and slightly outperform the
ML model on mean-based classification.

Presumably this is due to the difference in the number of trainable
weights. To test this, we define a 'constrained ML' network where
we subsample $NM$ weights to be trainable, and make sure those are
distributed approximately uniformly over the entries of $B^{(\!n\!)}$
for the different orders $n$. The remaining weights stay fixed at
their values at initialization. In comparison to the constrained ML
model, which has the same number of trainable parameters as the OSP,
we typically find a significant performance increase (see \prettyref{fig:OSP-vs-ML}
(d - f)).

\subsection{Identification of relevant orders from trained parameters\label{subsec:alphas}}

\begin{widetext}

\begin{figure}
\begin{centering}
\includegraphics[width=1\columnwidth]{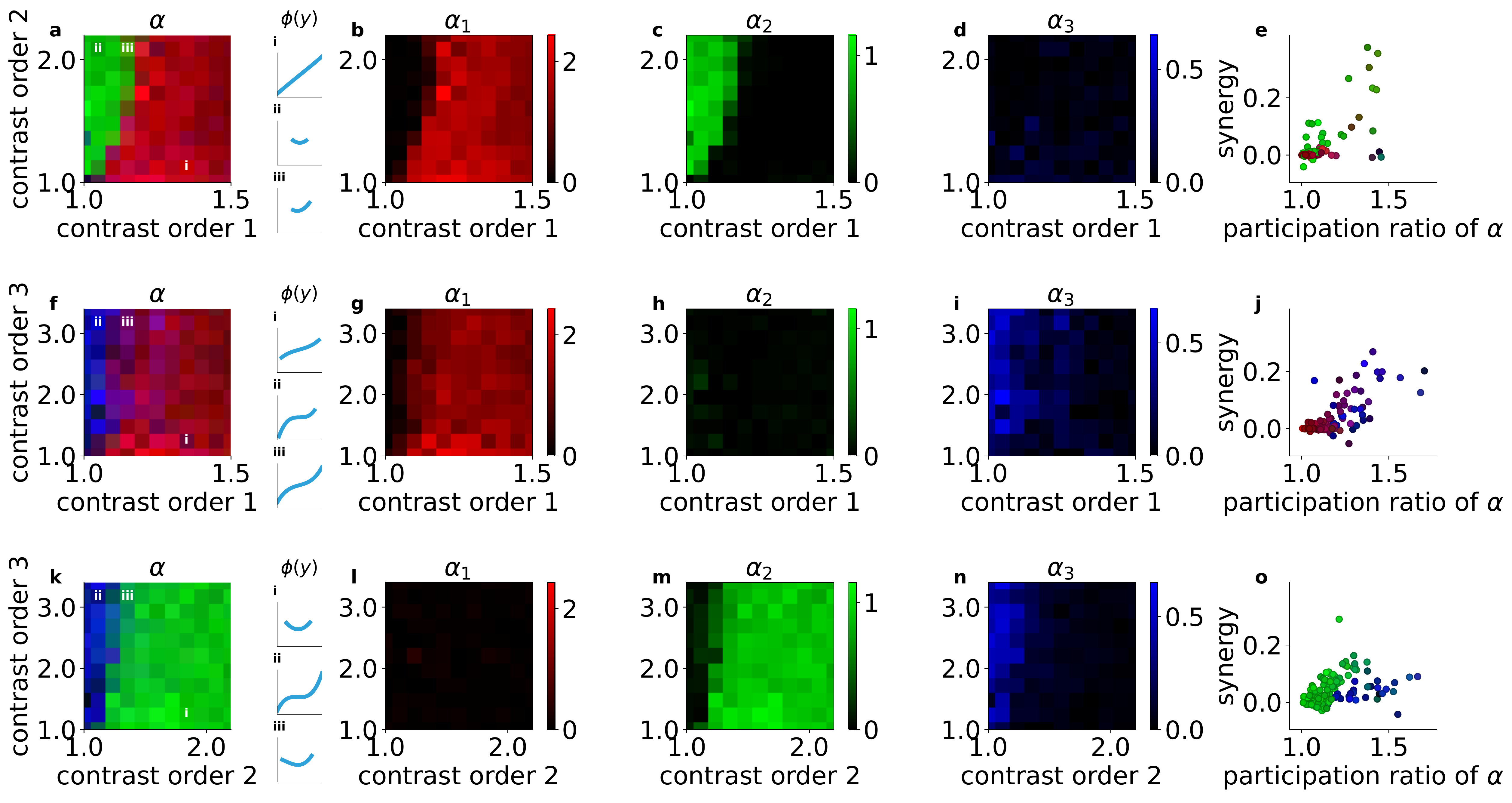}
\par\end{centering}
\caption{\textbf{Readout parameters for varying task difficulties.} Datasets
with two equal-sized classes which differ in two of their first three
cumulants (rows) are classified using the OSP. The weight vector $\alpha=(\alpha_{1},\alpha_{2},\alpha_{3})$
controlling the combination of orders one, two, and three is displayed
for the full model (\textbf{left} \textbf{column}) in color code,
where the value $\alpha_{n}$ for each order $n$ is displayed by
its corresponding rgb value: first order $\alpha_{1}$ (\textbf{red}),
second order $\alpha_{2}$ (\textbf{green}), and third order $\alpha_{3}$
(\textbf{blue}). Insets show the corresponding gain function at different
points of the parameter regime for the OSP. The second to forth column,
respectively, show the individual values of each $\alpha_{1\le i\le3}$.
The synergy is displayed over the participation ratio of $\alpha$
\textbf{(right column)}. On the axes of each plot the contrast between
the classes is varied, starting from one (no class difference) to
an arbitrarily chosen upper scale.\label{fig:Classification-alphas}}
\end{figure}

\end{widetext}

In the trained OSP, the $\alpha$ values describe how the different
input cumulants are combined in the readout to achieve classification;
see Methods for details, \prettyref{fig:setup-schematic}(b). All
parameters prior to the application of the non-linear gain function
are the same for each statistical order. One may therefore interpret
the absolute value of each component $|\alpha_{n}|$ as a measure
of the contribution of order $n$ to the classification. In \prettyref{fig:Classification-alphas}
each order $n$ is color-coded in red (mean), green (covariance) and
blue (third order), and the color value is proportional to $|\alpha_{n}|$,
scaled by the reciprocal of the largest $|\alpha_{n}|$ of all datasets.

As expected, the values $|\alpha_{n}|$ increase depending on the
input contrasts corresponding to cumulant order $n$. Again, we find
suppression in $\alpha$ due to the competition between the orders
when the contrast in one of the informative orders dominates over
the other. In contrast to the single order accuracy, however, we here
find overlapping regions of non-vanishing $\alpha_{n}$ corresponding
to the individual informative orders $n$. These are necessary to
form the synergy regions found in \prettyref{fig:Acc-synthetic}.
Thus, both informative orders are detected in this region. We find
again in \prettyref{fig:Classification-alphas} the effect of the
bias in \prettyref{sec:Training} that favors lower cumulant orders
over higher ones in the training, which is further amplified by the
normalization imposed on the $\Vert\alpha\Vert$.

In \prettyref{fig:Acc-synthetic}, we observed a synergy effect in
the performance along the boundary between regions where individual
cumulant orders dominate the classification. These regions coincide
with the regions where only one individual $\alpha_{n}$ is non-vanishing.
On the boundary between two such regions, both the corresponding $\alpha$
are non-vanishing. To confirm that the synergy is indeed linked to
this boundary, we display the synergy found in \prettyref{fig:Acc-synthetic}
to the participation ratio of $\alpha$ (\prettyref{eq:participation-ratio}),
which we use as a measure of sparsity here. Indeed, the measures are
correlated, although the participation ratio does not account for
the fact that $\alpha_{n}$ scales up to different maxima for each
order $n$.

We next investigate the corresponding nonlinear gain function $\phi$
shaped by the trained $\alpha$. Noticeably, higher order components
contribute significantly despite their lower combination parameter
$\alpha_{n}$. We display the gain functions after training for three
contrast combinations per synthetic dataset, see insets (i), (ii)
and (iii) in \prettyref{fig:Classification-alphas}. The regions are
dominated by a single order, so the gain functions are nearly ideal
linear, quadratic, or cubic functions, as expected. In the synergy
regions, more complex gain functions arise.

We would like to highlight that the identification of the optimal
gain function is specific to the network architecture of the OSP.
Although in the ML network, efforts can be made to make $\alpha$
scale similarly to the OSP by rescaling each order's weights $B^{\text{ML}(\!n\!)}$
appropriately, the resulting order combination weights $\alpha$ do
not correspond to a gain function. This would require joint intermediate
variables $y^{t}$ that would be the same for each order as they naturally
appear in the OSP due to the shared weights.

To summarize, the parameters $\alpha$ can be used to read out which
statistical order is dominant to distinguish between the classes.
When several $\alpha_{i}$ are non-zero, a synergy from combining
cumulants of different orders can be expected and the ``optimal''
gain function after training differs from pure linear, quadratic,
or cubic functions.

\subsection{Extraction of statistical patterns from real datasets\label{subsec:real-world-data}}

\begin{figure}
\begin{centering}
\includegraphics[width=1\columnwidth]{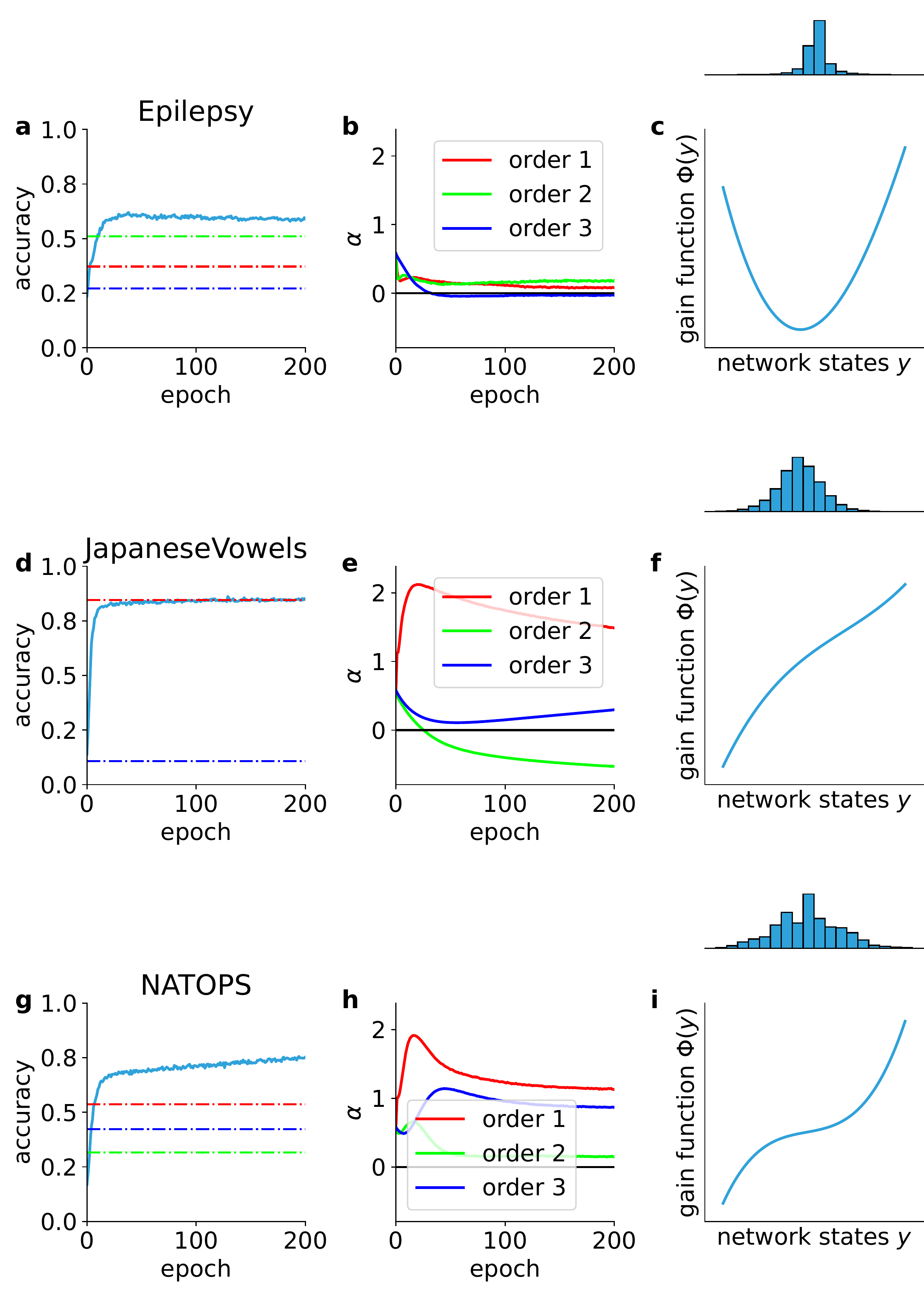}
\par\end{centering}
\caption{\textbf{Training on a few example benchmark datasets. }Training progress
and final gain function in the OSP for Epilepsy (a - c), JapaneseVowels
(d - f) and NATOPS (g - i) datasets, respectively. Accuracy of the
full network (blue curve) and single-order accuracy at the end of
training for orders one (red), two (green), and three (blue) as horizontal
lines (left). Evolution of cumulant combination weights $\alpha$
(\textbf{middle}). Gain function $\phi(y)$ resulting from final $\alpha$,
including the density of samples in the domain. \label{fig:Training-benchmark}}
\end{figure}

As a next step, we train and benchmark the OSP on $18$ datasets previously
used for benchmarking of time-series classification \citep{Chen15}.
They consist of multivariate time series with a diversity of input
dimensionalities, durations and sample sizes, as well as number of
classes. \prettyref{fig:Training-benchmark} displays the classification
results of the OSP for selected datasets that have different statistical
structures. For the example in \prettyref{fig:Training-benchmark}
(g-i), a large synergy from combining mostly the first and third order
evolves in the first few epochs. Decoding on individual statistical
orders, an accuracy of $53\,\%$ is achieved for the mean, $32\,\%$
for the covariance, and $42\,\%$ for the third order. The total classification
accuracy combining all orders reaches $75\,\%$. As we quantify synergy
by the difference between the final accuracy and the largest single
order accuracy, this corresponds to a synergy of $+22\,\%$.

The optimal gain function corresponding to this statistical structure
is consequently nearly point symmetric. The fluctuations of the samples
mostly pass through the region close to the origin. Only some classes
operate in the regime of strong non-linearity. \prettyref{fig:Training-benchmark}
(a-c) and \prettyref{fig:Training-benchmark} (d-f) show similar results
for the Epilepsy and JapaneseVowels dataset, respectively. Here, the
optimal gain functions are mostly quadratic or linear, respectively.

\begin{widetext}
\begin{center}
\begin{figure}
\begin{centering}
\includegraphics[width=1\columnwidth]{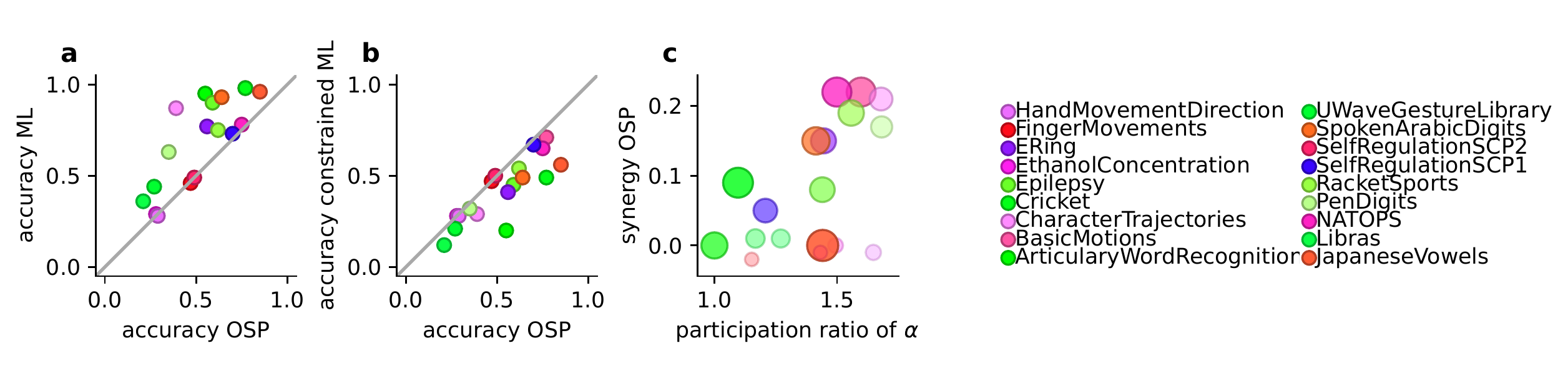}
\par\end{centering}
\caption{\textbf{Classification results on empirical datasets. (a)} Comparison
of the accuracy achieved in the ML and OSP models. Colors indicate
the datasets as listed on the side. \textbf{(b)} Analogous to (a)
with a constrained ML model with as many trainable parameters as the
OSP. \textbf{(c)} Synergy (see \prettyref{subsec:Training-process})
of the OSP for different datasets. Size and opacity indicate the improvement
in classification accuracy as compared to the pre-training accuracy
(larger size and more contrast for larger improvement). \label{fig:benchmark-comparison}}
\end{figure}
\par\end{center}

\end{widetext}

In \prettyref{fig:benchmark-comparison} (a), we compare the performance
of the benchmark datasets of the OSP to the ML model. Our goal is
less to compare to the state of the art with a perceptron-like model
(see \citep{Ruiz21} for recent results), but to showcase how appropriate
processing of statistical information can improve network capabilities.
Nevertheless, the performance of the OSP often approaches that of
the corresponding machine learning model. In most of the cases, both
OSP and ML perform above chance level, which we empirically evaluate
using similar architectures with untrained parameters. Accuracy remains
at chance level only for two datasets: HeadMovementDirection and SelfRegulationSCP2.

For synthetic data (\prettyref{fig:OSP-vs-ML}), the OSP yielded mostly
lower classification accuracy compared to the ML approach, but outperformed
a constrained ML model in which the number of trainable parameters
equals those of the OSP. We observe the same for the benchmark datasets
in \prettyref{fig:benchmark-comparison}: In comparison to the constrained
ML model, we typically find a significant performance increase (see
\prettyref{fig:benchmark-comparison} (b)), whereas the ML model that
trains directly on the full input cumulants exhibits even higher performance
(see \prettyref{fig:benchmark-comparison} (a)).

We furthermore extract also the relevant statistical orders in the
data by inspecting the trained values of $\alpha$. With the coloring
of each dataset according to its cumulant combination weights $\alpha=(\alpha_{1},\alpha_{2},\alpha_{3})=(\text{red},\text{green},\text{blue})$,
the broad color spectrum displayed in \prettyref{fig:benchmark-comparison}
shows a wide variety for the relevant statistical orders in the different
datasets.

In \prettyref{fig:benchmark-comparison} (c), we display the accuracy
increase with respect to the best single order for the different datasets.
While the color indicates the dataset, both size and transparency
encode the training performance of the OSP in relation to the pre-training
accuracy. Datasets that are less well classified by the OSP are thus
displayed smaller and more transparent. Based on the synthetic data,
we expect larger synergy where different orders are combined more
strongly. In \prettyref{fig:benchmark-comparison} (c), we therefore
relate synergy to the participation ratio $\rho$ (see \prettyref{subsec:Training-process}).
Although the results on synthetic data suggest a non-uniform weighting
between the different orders $n$, we observe synergy effects also
in complex datasets, specifically from combining more than one statistical
order. In this figure, we observe a correlation between the participation
ratio and the observed synergy in the different datasets.

\section{Discussion}

In this work, we analyzed statistical processing for time series by
a perceptron-like model with a tuneable non-linearity. We showed that
the form of the non-linearity determines which statistical patterns,
quantified via cumulants up to the third order, are transformed to
shape the network output, and how the different orders jointly contribute
to the output to solve the task. The gain function that is optimal
for the time series classification tasks depends sensitively on the
relative importance of class-specific differences (or contrasts) in
each cumulant. With the order selective perceptron, informative statistical
properties of the data can be revealed with minimal model complexity.

This minimal complexity is achieved by avoiding the explicit computation
of cumulants at several orders. Instead, the selection of the relevant
cumulants of the input $x^{t}$, via those of the intermediate variable
$y^{t}$, results from the combination of learning rules for the input
weights and the nonlinearity. While the input cumulants, in particular
the higher orders, require huge tensors for $N$-dimensional multivariate
time series with large $N$, the cumulants of $y^{t}$ scale polynomially
with $M$, the number of classes, which is already often lower. Since
additionally, only the diagonal of the network state cumulant is computed,
the computational cost reduces further to only $M$ entries per order.
Furthermore, the number of trainable parameters reduces from $K+M\cdot(N+N^{2}+\ldots+N^{K})$
for the ML network to $K+M\cdot N$ for the OSP (with $K$ the maximum
cumulant order).

Naturally, this reduced model complexity comes at lower performance,
as compared to the explicit computation of all cumulant orders. However,
the network layer $B$ practically accumulates as much ``information''
as possible from the input cumulants in the intermediate variables
for classification: We have showed that the OSP performs in between
the ML model acting on the same cumulants as the OSP and the constrained
ML network, which trains only a reduced set of weights of the original
ML architecture (to match the trained parameters of the OSP). The
original ML network thereby acts as a theoretical upper bound for
the OSP, as it can freely combine all of the input cumulants. The
fact that the OSP outperforms the constrained ML model shows the efficient
computation in the OSP despite the competition between the different
orders. For neural classifiers that perform at the current state of
the art, this may suggest as a metric their flexibility with regard
to the adaptive propagation of cumulants. For example, in the spirit
of \citep{lawrie21covariance} which have showed superior performance
of covariance encoding in reservoir computing compared to linear encoding,
an order selective perceptron could be combined with a recurrent reservoir
to combine the benefits of both. A thorough study of more sophisticated,
large-scale architectures that utilize high order cumulants in this
regard remains for future work.

The regularization-induced competition between the orders in fact
strongly affects how the OSP combines cumulants for classification.
We created synthetic data with tuneable cumulant structure to investigate
this dependence in detail. Our SDE-based algorithm generates data
with known cumulants up to third order. Field theory can be used to
obtain both estimates of the higher order structures and more accurate
estimates of the cumulants, which are given by a series of coefficients
with decreasing weights. It is, however, limited to statistics that
does not deviate too strongly from the exactly solveable Gaussian
theory. To the best of our knowledge, there has to date been no algorithm
that would provide the desired controlled data for multivariate time
series. It is noteworthy, though, that there exists the CuBIC technique
\citep{Staude10_16_4,Staude10_327}, which creates spike trains with
a given a desired cumulant structure. The algorithm presented here
can also be used to create static stimuli.

Using such synthetic data with known underlying statistical structure,
we found that the OSP preferentially classifies using only a single
order, relying on only one cumulant, if the difference between the
classes is more expressed in this cumulant than in the others. This
is reinforced by the regularization of the cumulant combining layer,
although the same behavior qualitatively also occurs for unconstrained
$\alpha$. In that case, however, the network is classifying a bit
less well overall. On the boundary between two different preferred
orders, we observe synergy effects from combining cumulants through
different paths. The class contrast where this boundary occurs appears
however non-trivial. From both the position of the boundary and the
amplitudes of the order selection parameter $\alpha$ in the regions
next to the boundary, we conclude that the network tends to prefer
lower order cumulants over higher ones, and selects the preferred
order using this hierarchy. Applied to real-world applications that
serve as benchmark datasets, often several orders are combined, and
we regularly observe synergistic effects. Combination of statistical
information appears all the more important for complicated data structures.
For any combination of orders, there is a dataset found among them
whose order selection parameters filter for just this combination.
The gain functions that the order combination parameters translate
to are consequently uniquely shaped.

Neuronal network architectures similar to the OSP have been widely
used for different purposes in the context of machine learning \citep{can2021emergence}
and computational neuroscience \citep{Sompolinsky88_259,Schuecker18_041029,Engelken22_09916},
with nonlinear neurons governed by bounded sigmoid-like profiles like
in the Wilson-Cowan-model \citep{Wilson72_1,Wilson73}, or rectified
linear units (ReLU) \citep{nair10}. \citep{rosenzweig20} give a
thorough review of different gain functions and their influence on
computational performance. We stress that our study differs from the
use of fixed (non-trainable) gain functions that are typically used.
The choice of the gain function, however, matters for the the statistical
processing performed by the network. The purely linear gain function
corresponds to the original perceptron \citep{Rosenblatt58,Minsky69,Widrow60_96}
where each input cumulant is mapped to its counterpart at the same
order in the output. In contrast, a nonlinear readout can perform
cross-talk between input cumulants of several orders, combining them
into e.g. the output mean. On the other hand, correlation patterns
have recently been proposed as the basis of ``information'' embedded
in time series that can be processed for e.g. classification \citep{Gilson20_1}.
This so-called covariance (de)coding has been shown to yield larger
pattern capacity than the classical perceptron that relies on mean
patterns when applied to time series \citep{Dahmen20_354002}.

In conclusion, the choice of the gain function at the same time is
a choice on what statistical information the network should be sensitive
to. Gain functions that are more complex than simple polynomials thus
combine many, if not all, orders of cumulants. We chose a polynomial
here for interpretability, in practice high performance could be achieved
by other choices of tuneable non-linearities. However, a fixed gain
function means also that there is a fixed relation between the different
statistical contributions of the underlying probability density. Translating
a given gain function into a Taylor series around a working point
can give insight into what cumulants a network may be particularly
sensitive to, although this does not yield a one-to-one translation
to our adaptive gain function. The here employed de-meaning of the
time series translates to a different weighting of cumulants depending
on the sample's mean; it still holds that this form of input processing
is fixed and pre-defined. In biological data, adaptive gain has been
observed \citep{diaz2008intrinsic,maravall2013transformation}, although
not identical to the simple mechanism presented here.

We therefore suggest making the choice of gain function with care.
There have been, in fact, works on comparing how well different prominent
gain functions work in machine learning contexts \citep{nwankpa2018activation,szandala2021review},
and some commonly used gain functions include trainable parameters
\citep{lau2018review}. A fully trainable gain function is presented
in \citep{ertuugrul2018novel}, which is of a similar polynomial type
as presented here. The aim of these works is to provide optimal performance
for sophisticated networks on complicated tasks. We offer interpretability
of the gain function in terms of statistical processing and provide
a link to biological neural networks.

The gain variability found for example by \citep{diaz2008intrinsic,maravall2013transformation}
shows that neurons may be sensitive to the statistical features of
their inputs, and the variability of activity also appears to be linked
to behavior \citep{Riehle97_1950,Kilavik09_12653,Shahidi19}. It may
also play a role in representing uncertainty in terms of probabilistic
inference \citep{berkes2011spontaneous,orban2016neural,henaff2020representation,festa2021neuronal}.
Structured variability thus seems to play an integral role in biological
neural networks. As far as information encoding through neural correlations
is superior to mean-based representations, this may help understand
why such sensitivity has emerged in neuronal circuits. This reasoning
has already inspired the covariance perceptron \citep{Gilson19_562546,Dahmen20_354002}.

Of course, just like the covariance perceptron, the order selective
perceptron is not a biologically detailed network model. It is built
upon the most fundamental building block, a feedforward layer of neurons,
followed by a gain function. Instead of individual neuron's spiking
activity, it treats populations of neurons as a unit with an activity
that resembles the average firing rate. The linear summation of inputs
resembles closely the neuron dynamics typically assumed for cortical
models \citep{Sompolinsky88_259,Amari72_643,rajan10_011903,Hermann12_018702,kadmon15},
but also in machine learning, including the perceptron \citep{Rosenblatt58,Minsky69,Widrow60_96}.
The main difference to previously presented methods is the assumption
of information being hidden in the higher order statistics of the
network activity. While the propagation of stimuli through neural
networks is widely accepted to be modeled suitably by linear summation
of inputs and application of subsequent transfer functions, this point
of view draws more towards understanding the computation based on
the inputs.

Both regarding computational capacity and biological realism, it would
be interesting to study also a multi-layer or a recurrently connected
version of the order selective perceptron, which extends outside the
scope of this work. It is of particular relevance to study the performance
gain from time-lagged cumulants. Further, the cross-correlations of
the intermediate network state, which are currently not influencing
the classification, may be useful in future work.
\begin{acknowledgments}
This work was partly supported by European Union Horizon 2020 grant
945539 (Human Brain Project SGA3), the Helmholtz Association Initiative,
the Networking Fund under project number SO-092 (Advanced Computing
Architectures, ACA), the BMBF Grant 01IS19077A (Juelich), and the
Excellence Initiative of the German federal and state governments
(G:(DE-82)EXS-PF-JARASDS005).

\vfill{}
\newpage{}
\end{acknowledgments}

\appendix

\section{Computation of cumulants for the artificial data model\label{sec:Compute-statistics}}

The cumulants of the probability density $p(x)\sim\exp\left(-\beta\left(m^{T}x+\frac{1}{2}x^{T}Jx+\frac{1}{3!}\sum_{ijk}K_{ijk}x_{i}x_{j}x_{k}\right)\right)$
(with a potential barrier) can be derived using field theory \citep{Helias19_10416}.
To this end, the Gaussian model with $p(x)\sim\exp\left(-\beta\left(m^{T}x+\frac{1}{2}x^{T}Jx\right)\right)$
acts as a baseline, and the cubic part forms a small correction for
states near those expected from the Gaussian distribution. From the
Gaussian baseline follows, that the propagator of the system is its
covariance,
\begin{equation}
\Sigma_{ij}=\frac{1}{\beta}(J^{-1})_{ij}\,.
\end{equation}
With $m$ acting as a source term, the mean follows as
\begin{equation}
\mu_{i}=-\sum_{j}(J^{-1})_{ij}m_{j}\,.
\end{equation}
The cubic correction leads to a three-point vertex, which has the
value $-\frac{\beta}{3!}K$. In first order corrections to the Gaussian
theory, therefore, diagrams with a single vertex need to be computed.
For the mean, this leads to a tadpole diagram, where two of the three
legs of a vertex are connected by the propagator, and a single external
leg. There are three different such diagrams from the three possible
choices of the external legs, leading to the corrected mean
\begin{equation}
\mu_{i}=-\sum_{j}(J^{-1})_{ij}m_{j}-\frac{1}{2\beta}\sum_{jkl}(J^{-1})_{ij}K_{jkl}(J^{-1})_{kl}\,.
\end{equation}
This mean may not deviate too strongly from the Gaussian one. There
are no diagrams with a single three-point vertex and two external
legs, so up to first order, the covariance stays as in the Gaussian
case. The third order cumulant is simply a vertex with a propagator
on each of its legs. A prefactor of $3!$ needs to be included for
those. To summarize, this leads to the cumulants
\begin{align}
\mu_{i} & \approx-\sum_{j}(J^{-1})_{ij}m_{j}-\frac{1}{2\beta}\sum_{jkl}(J^{-1})_{ij}K_{jkl}(J^{-1})_{kl}\,,\\
\Sigma_{ij} & \approx\frac{1}{\beta}(J^{-1})_{ij}\,,\\
S_{ijk} & \approx-\frac{1}{\beta^{2}}\sum_{i^{\prime}j^{\prime}k^{\prime}}K_{i^{\prime}j^{\prime}k^{\prime}}(J^{-1})_{ii^{\prime}}(J^{-1})_{jj^{\prime}}(J^{-1})_{kk^{\prime}}\,.
\end{align}

\section{How to choose suitable parameters\label{sec:Select-K}}

In order to obtain reasonable results, the parameters need to be chosen
with care. The cubic part $K$ may not be too large, to ensure the
system is both not too unstable and close enough to the Gaussian case
that we can use field theory to approximate the cumulant statistics,
however large enough that substantial third order statistics arises
for our desired dataset. To this end, we compare the expected fluctuations
of states with the width of the safe part of the potential, the region
between the local minimum and the local maximum.

For this comparison to be useful, however, we first need to ensure
that during all times of the evolution of the SDE, the data stays
close to the Gaussian theory. Therefore, we simulate the SDE with
random initial conditions drawn from a Gaussian distribution with
mean and covariance identical to those expected according to \prettyref{sec:Compute-statistics}.
Furthermore, we set the source $m=0$ to simplify calculations. The
mean of the data can anyway be adjusted by a constant shift $b$ of
all data points after simulating the SDE. The quadratic part $J$
needs to be chosen as a positive definite matrix that fits the desired
covariance.

The remaining parameter to determine is therefore only $K$. It has
to be chosen such that for $m=0$ and a given $J$, there is no direction
of the potential where the Lagrangian $L[x]=\frac{1}{2}x^{T}Jx+\frac{1}{3!}\sum_{ijk}K_{ijk}x_{i}x_{j}x_{k}$
rises monotonically. For a start, we'll assume that the cubic part
will be of rank one, in the sense that there is a single direction
in state space where the Lagrangian is cubic, and in all directions
orthogonal to this one, the potential remains quadratic. We'll be
then able to compose stronger cubic potentials.

With $x^{\prime}$ in direction of non-vanishing cubic interactions,
this means that there must be real solutions to the necessary conditions
for local minima or maxima, 
\begin{equation}
\sum_{j}J_{ij}x_{j}^{\prime}+\frac{1}{2}\sum_{jk}K_{ijk}x_{j}^{\prime}x_{k}^{\prime}=0\,.
\end{equation}
Only then, a stable region will exist for the data to evolve around.
With $K$ being rank one, it can be composed as an outer product of
vectors $v$ as 
\begin{align}
K_{ijk} & =v_{i}v_{j}v_{k}\,.
\end{align}
The condition for local extrema becomes
\begin{equation}
Jx^{\prime}+\frac{1}{2}v(v^{T}x{}^{\prime})^{2}=0\,.
\end{equation}
As desired, in any direction orthogonal to $v$ (i.e. where $v^{T}x{}^{\prime}=0$),
then the potential is quadratic and thereby stable. It therefore suffices
to look at $x^{\prime}\parallel v$. In that case, when $x^{\prime}=\Vert x\Vert e^{(\!v\!)}$,
\begin{align}
Jx^{\prime}+\frac{1}{2}v(v^{T}x{}^{\prime})^{2} & =\Vert x\Vert Je^{(\!v\!)}+\frac{1}{2}\Vert x\Vert^{2}\,\Vert v\Vert^{3}\:e^{(\!v\!)}\\
 & =0
\end{align}
where $e^{(\!v\!)}$ denotes the unit vector in direction of $v$.
If $e^{(\!v\!)}$ is an eigendirection of $J$ with eigenvalue $\lambda_{(\!v\!)}$,
this simplifies further. Then, one can solve for $\Vert v\Vert$ to
obtain $v=\Vert v\Vert\:e^{(\!v\!)}=-2^{\frac{1}{3}}\Vert x\Vert^{-\frac{1}{3}}\lambda_{(\!v\!)}^{\frac{1}{3}}\:e^{(\!v\!)}$.
Now, the extrema of the potential lie on a line in direction $e^{(\!v\!)}$,
at $x^{\prime}=0$ and $x^{\prime}=\Vert x\Vert e^{(\!v\!)}$. The
distance between the minimum and maximum therefore is $\Vert x\Vert$
and can be chosen to determine the cubic part of the Lagrangian. We
can here choose to set
\begin{equation}
\Vert x\Vert^{2}=s^{2}e^{(\!v\!)\mathrm{T}}\Sigma e^{(\!v\!)}
\end{equation}
such that we expect $s$ standard deviations to fit in this stable
part of the potential (see \prettyref{sec:Compute-statistics}).

We can lastly compare the third order cumulant to the second to find
values of $s$ for the third order statistics to be strong enough
compared to the fluctuations. To this end, we can map the second and
third order statistics in some eigendirection $e^{(\!v^{\prime}\!)}$
of $J$ (which is also an eigendirection of $\Sigma$ with eigenvalue
$(\beta\lambda_{v^{\prime}})^{-1}$) to find
\begin{equation}
\left(e^{(\!v^{\prime}\!)\mathrm{T}}\Sigma e^{(\!v^{\prime}\!)}\right)^{\frac{3}{2}}=\left(\beta\lambda_{(\!v^{\prime}\!)}\right)^{-\frac{3}{2}}
\end{equation}
and
\begin{align}
\sum_{ijk}S_{ijk}e_{i}^{(\!v^{\prime}\!)}e_{j}^{(\!v^{\prime}\!)}e_{k}^{(\!v^{\prime}\!)} & =-\frac{1}{\beta^{2}}\left(e^{(\!v^{\prime}\!)\mathrm{T}}J^{-1}v\right)^{3}\nonumber \\
 & =-\frac{1}{\beta^{2}}\left(e^{(\!v^{\prime}\!)}\lambda_{(\!v\!)}^{-1}v\right)^{3}\nonumber \\
 & =-\frac{\Vert v\Vert^{3}}{\beta^{2}\lambda_{(\!v\!)}^{3}}\delta_{vv^{\prime}}\nonumber \\
 & =\frac{1}{\Vert x\Vert\beta^{2}\lambda_{(\!v\!)}^{2}}\delta_{vv^{\prime}}\nonumber \\
 & =\frac{1}{s\left(e^{(\!v\!)\mathrm{T}}\Sigma e^{(\!v\!)}\right)^{\frac{1}{2}}\beta^{2}\lambda_{(\!v\!)}^{2}}\delta_{vv^{\prime}}\nonumber \\
 & =\frac{1}{s}\left(\beta\lambda_{(\!v\!)}\right)^{-\frac{3}{2}}\delta_{vv^{\prime}}.
\end{align}
From this follows 
\begin{equation}
\left|\frac{\sum_{ijk}S_{ijk}e_{i}^{(\!v\!)}e_{j}^{(\!v\!)}e_{k}^{(\!v\!)}}{\left(\sum_{ij}\Sigma_{ij}e_{i}^{(\!v\!)}e_{j}^{(\!v\!)}\right)^{\frac{3}{2}}}\right|=\frac{1}{s}
\end{equation}
for $e^{(\!v\!)}$ in direction of the cubic perturbation.

It is straightforward to then construct a Lagrangian with a safe cubic
potential in any direction by summing over all eigendirections of
$J$,
\begin{align}
K_{ijk} & =\sum_{(\!v\!)}v_{i}^{(\!v\!)}v_{j}^{(\!v\!)}v_{k}^{(\!v\!)}\,,\\
v^{(\!v\!)} & =-s_{(\!v\!)}^{-\frac{1}{3}}\left(e^{(\!v\!)T}\Sigma e^{(\!v\!)}\right)^{-\frac{1}{6}}\lambda_{(\!v\!)}^{\frac{1}{3}}\:e^{(\!v\!)}\,,
\end{align}
where $s_{(\!v\!)}$ controls the strength of the cubic potential
in each eigendirection of $J$. For or purposes, $s=s_{(\!v\!)}=3$
was used in all directions.

\section{Comparison to constrained order perceptron models\label{sec:other-synergy}}

\begin{widetext}

\begin{figure}
\begin{centering}
\includegraphics[width=1\columnwidth]{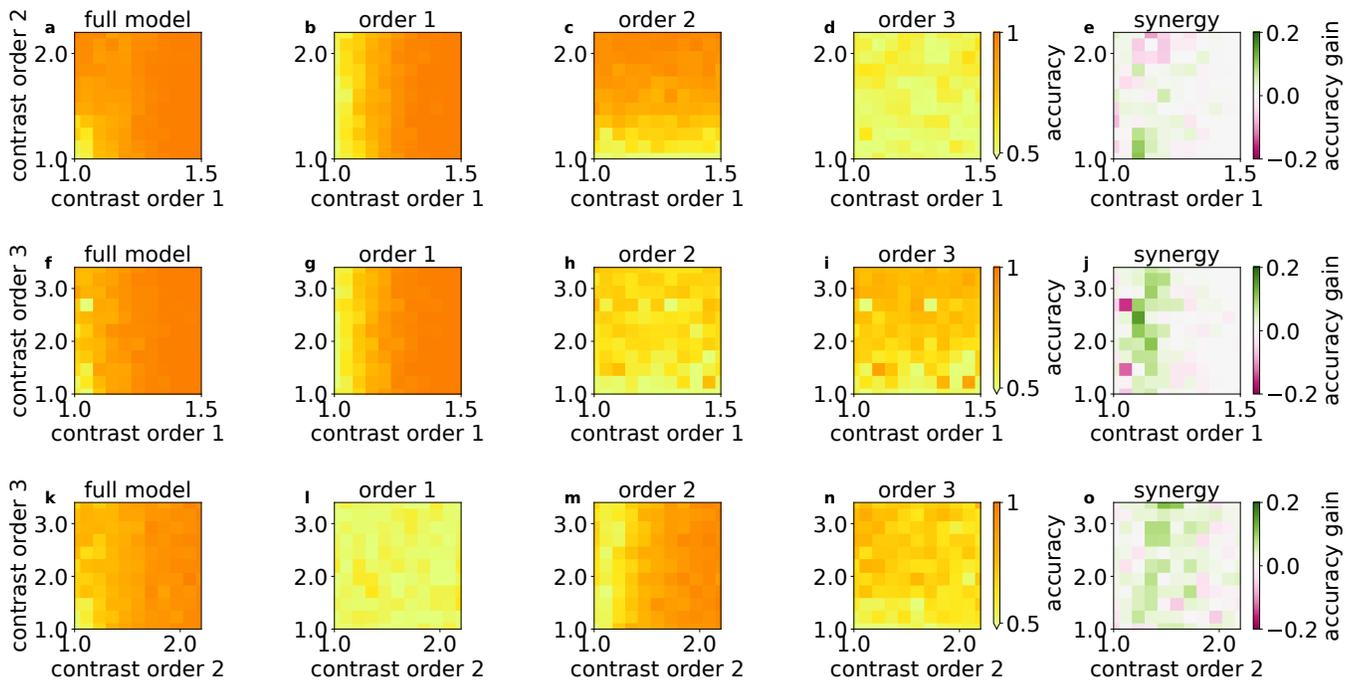}
\par\end{centering}
\caption{Performance comparison of the OSP and a constrained OSP. (a) Accuracy
of the OSP as in \prettyref{fig:Acc-synthetic}. (b) Accuracy of a
constrained OSP with trainable $\alpha_{1}$ and $\alpha_{2,3}=0$.
(c, d) Analogous to (b) with $\alpha_{2,3}$ trainable, respectively.
(e) Accuracy gain of the OSP over the constrained OSP. In (a - e),
classes are separated by a difference in the mean and covariance.
In (f - j), classes are separated by a difference in the mean and
third order cumulant. In (k - o), classes are separated by a difference
in the covariance and third order cumulant.\label{fig:alternative-synergy}}
\end{figure}

\end{widetext}

We defined synergy as the performance gain of the OSP compared to
a pruned OSP, where all except a single entry of $\alpha$ is set
to zero. This is motivated by the ability of $\alpha$ to indicate
the contribution of corresponding statistical orders to the classification
decision, as deduced from the competition between its entries. A different
angle to the question of how much the OSP gains can be whether and
when the OSP outperforms a constrained OSP, in which entries of $\alpha$
are set to zero throughout training.

\prettyref{fig:alternative-synergy} shows a comparison of the performance
of these models. Constrained OSP ignore the contrasts in cumulants
that they are not sensitive to and perform well above a minimum contrast
in their corresponding cumulant order. Without the competition imposed
on the unconstrained OSP, the performance is typically higher than
the single order accuracy given by the pruned networks in \prettyref{fig:Acc-synthetic}.
Consequently, the difference of accuracy between these models is lower.
However, we still observe a tendency towards an accuracy gain along
an area in the space of contrasts. Noticeably, it coincides mostly
with the border between areas in which individual orders dominate
the unconstrained OSP (see \prettyref{fig:Acc-synthetic}). The OSP
therefore actively benefits from combining different cumulants for
classification.

\section{Synthetic data, ML model\label{sec:Artificial-ML}}

Training of the ML network on the synthetic data is displayed in \prettyref{fig:Classification-alphas-ML}.
In \prettyref{fig:Classification-alphas-constrained}, the training
is repeated with an ML model, of which only $MN$ weights are randomly
selected to be trainable.

\begin{widetext}

\begin{figure}
\begin{centering}
\includegraphics[width=1\columnwidth]{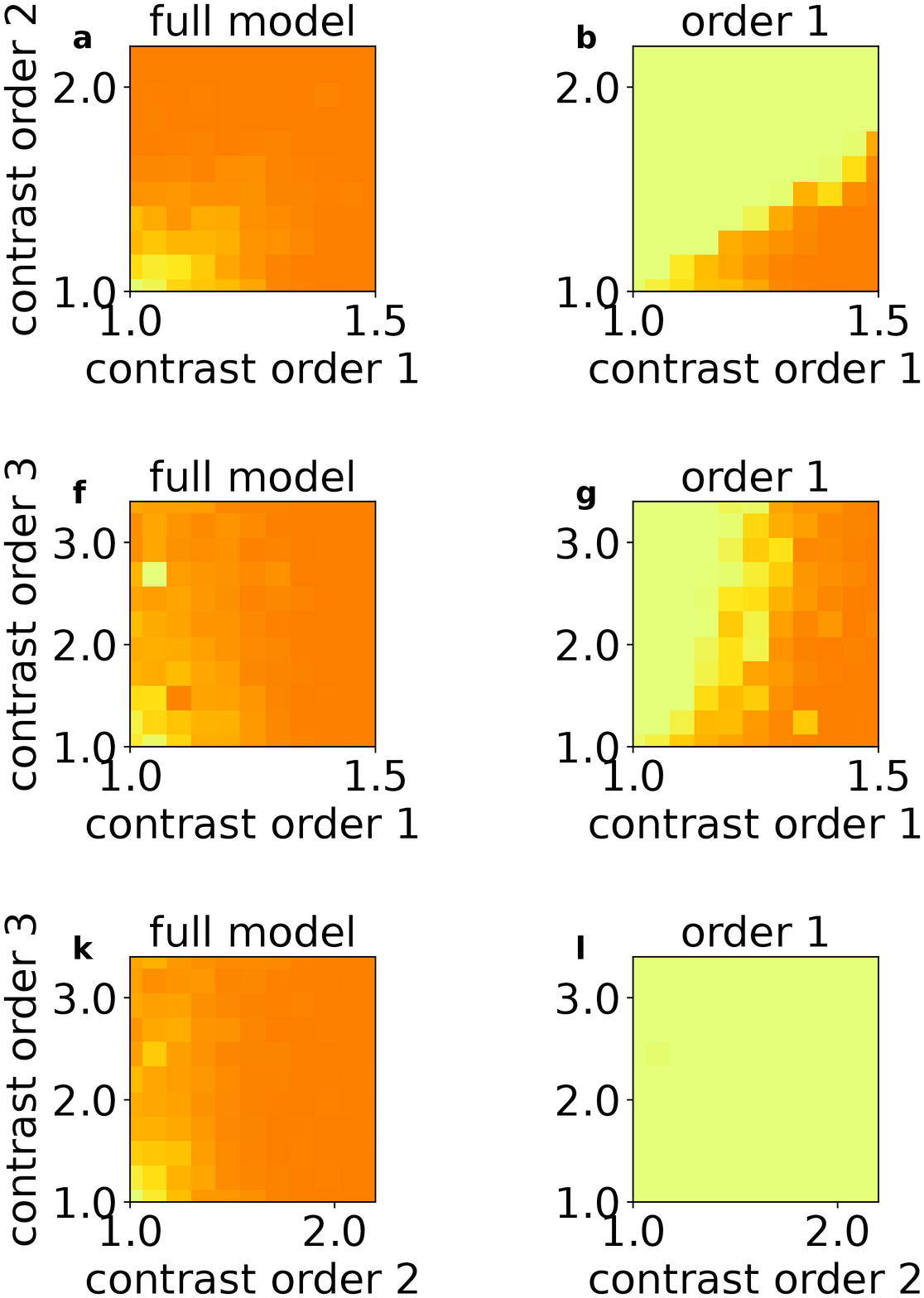}
\par\end{centering}
\caption{\textbf{Readout parameters for the} \textbf{ML network with varying
task difficulties.} Datasets with two equal-sized classes which differ
in two of their first three cumulants (\textbf{rows}) are classified
using the ML network. The cumulant combination weight $\alpha$ is
displayed for the full model (\textbf{left}) in color code, for the
first (\textbf{red}), second (\textbf{green}) and third order (\textbf{blue}).
Insets show the corresponding gain function at different points of
the parameter regime for the OSP. Consecutively, starting from the
second from left, $\alpha_{i}$ for each individual order $i$ from
one (mean) to three (third order cumulant) is showed. The axes of
each diagram display the contrast between the classes underlying the
datasets, starting from one (no class difference) to an arbitrarily
chosen upper scale. In (a - e), classes are separated by a difference
in the mean and covariance. In (f - j), classes are separated by a
difference in the mean and third order cumulant. In (k - o), classes
are separated by a difference in the covariance and third order cumulant.\label{fig:Classification-alphas-ML}}
\end{figure}

\begin{figure}
\begin{centering}
\includegraphics[width=1\columnwidth]{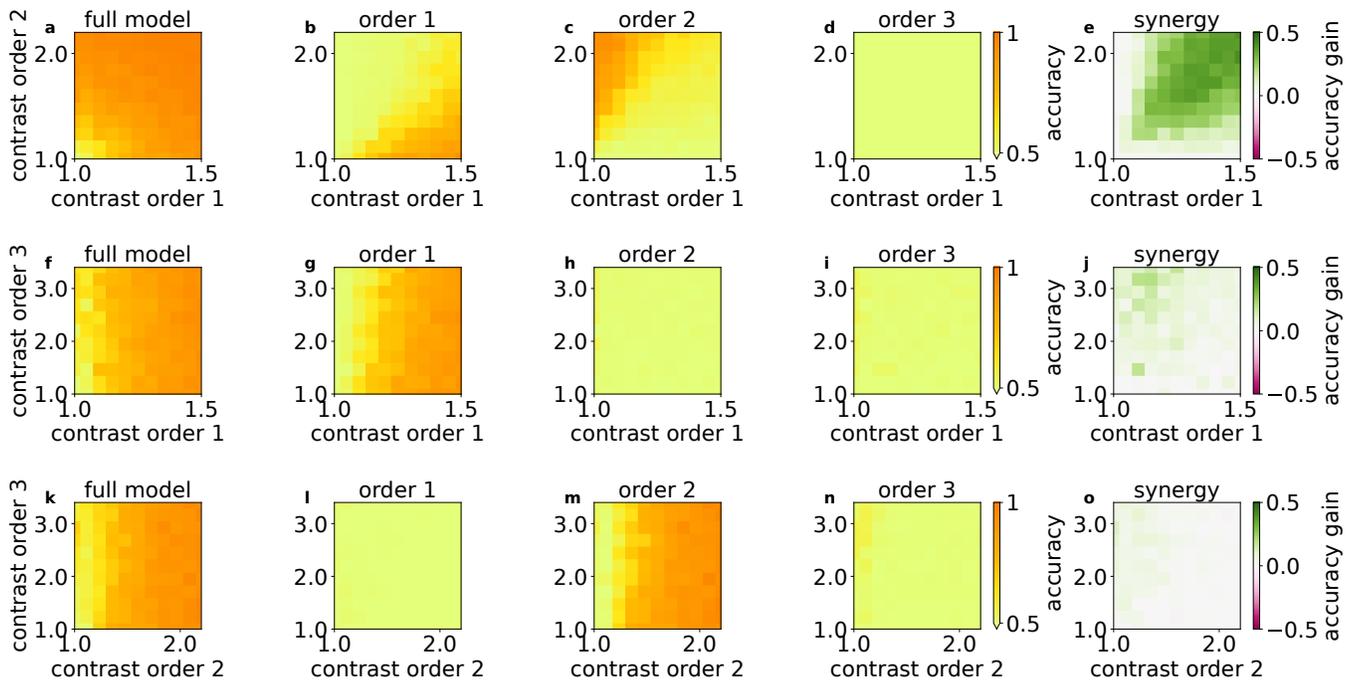}
\par\end{centering}
\caption{\textbf{Readout parameters for the constrained ML model with varying
task difficulties.} Datasets with two equal-sized classes which differ
in two of their first three cumulants (\textbf{rows}) are classified
using the constrained ML network. The cumulant combination weight
$\alpha$ is displayed for the full model (\textbf{left}) in color
code, for the first (\textbf{red}), second (\textbf{green}), and third
order (\textbf{blue}). Insets show the corresponding gain function
at different points of the parameter regime for the OSP. Consecutively,
starting from the second from left, $\alpha_{i}$ for each individual
order $i$ from one (mean) to three (third order cumulant) is showed.
The axes of each diagram display the contrast between the classes
underlying the datasets, starting from one (no class difference) to
an arbitrarily chosen upper scale. In (a - e), classes are separated
by a difference in the mean and covariance. In (f - j), classes are
separated by a difference in the mean and third order cumulant. In
(k - o), classes are separated by a difference in the covariance and
third order cumulant.\label{fig:Classification-alphas-constrained}}
\end{figure}

\end{widetext}\vfill{}
\newpage{}

\bibliographystyle{plainnat}
\bibliography{brain}

\end{document}